\documentclass{emulateapj}

\bibpunct{(}{)}{;}{a}{}{,}

\slugcomment{To appear in ApJ}

\shorttitle{ACBAR Power Spectrum}
\shortauthors{Reichardt et al.}

\usepackage{amsmath}
\usepackage{amssymb}

\begin{document}

\title{High resolution CMB power spectrum from the complete ACBAR data set.}


\author{ C.L. Reichardt\altaffilmark{1},
P.A.R. Ade\altaffilmark{2},
J.J. Bock\altaffilmark{1,3}, 
J.R. Bond\altaffilmark{4},
J.A. Brevik\altaffilmark{1},
C.R. Contaldi\altaffilmark{5},
M.D. Daub \altaffilmark{6}, 
J.T. Dempsey \altaffilmark{7},
J.H. Goldstein\altaffilmark{8,9}, 
W.L. Holzapfel \altaffilmark{6}, 
C.L. Kuo\altaffilmark{1,10},
A.E. Lange\altaffilmark{1,3}, 
M. Lueker\altaffilmark{6},
M. Newcomb\altaffilmark{11},
J.B. Peterson\altaffilmark{12},
J. Ruhl\altaffilmark{8}, 
M.C. Runyan\altaffilmark{1},
Z. Staniszewski\altaffilmark{8}}

\altaffiltext{1}{Observational Cosmology, California Institute of Technology, MS 59-33, Pasadena, CA 91125}
\altaffiltext{2}{Department of Physics and Astronomy, Cardiff University, CF24 3YB Wales, UK}
\altaffiltext{3}{Jet Propulsion Laboratory, 4800 Oak Grove Drive, Pasadena, CA 91109}
\altaffiltext{4}{Canadian Institute of Theoretical Astrophysics, University of Toronto, Toronto, Ontario, M5S3H8,
Canada}
\altaffiltext{5}{Blackett Laboratory, Imperial College, Prince Consort Road, London, SW7 2AZ, U.K.}
\altaffiltext{6}{Department of Physics, University of California, Berkeley, CA 94720}
\altaffiltext{7}{Joint Astronomy Centre, Hilo HI 96720}
\altaffiltext{8}{Department of Physics, Case Western Reserve University, Cleveland, OH 44106}
\altaffiltext{9}{Aret\'{e} Associates, Arlington, VA 22202}
\altaffiltext{10}{Department of Physics and KIPAC, Stanford University, Stanford, CA 94305}
\altaffiltext{11}{Yerkes Observatory, 373 W. Geneva Street, Williams Bay, WI 53191}
\altaffiltext{12}{Department of Physics, Carnegie Mellon University, Pittsburgh, PA 15213}

\begin{abstract}

  In this paper, we present results from the complete set of cosmic
  microwave background (CMB) radiation temperature anisotropy
  observations made with the Arcminute Cosmology Bolometer Array
  Receiver (ACBAR) operating at $150\,$GHz.  We include new data from
  the final 2005 observing season, expanding the number of
  detector-hours by 210\% and the sky coverage by 490\% over that used
  for the previous ACBAR release.  As a result, the band-power
  uncertainties have been reduced by more than a factor of two on
  angular scales encompassing the third to fifth acoustic peaks as
  well as the damping tail of the CMB power spectrum.  The calibration
  uncertainty has been reduced from 6\% to 2.1\% in temperature
  through a direct comparison of the CMB anisotropy measured by ACBAR
  with that of the dipole-calibrated WMAP5 experiment.  The measured
  power spectrum is consistent with a spatially flat, $\Lambda$CDM
  cosmological model. We include the effects of weak lensing in the power spectrum model computations and find that this significantly improves the 
fits of the models to the combined ACBAR+WMAP5 power spectrum.
The preferred strength of the lensing is consistent with theoretical expectations. On fine angular scales, there is weak evidence ($1.1\sigma$) for excess power above the level expected from primary anisotropies. We expect any excess power to be dominated by the combination of emission from  
dusty protogalaxies and the Sunyaev-Zel'dovich effect (SZE).
However, the excess observed by ACBAR is significantly smaller than the excess power at $\ell >2000$ reported by the CBI experiment operating at $30\,$GHz. 
Therefore, while it is unlikely that the CBI excess has a primordial origin;
the combined ACBAR and CBI results are consistent with the source of
the CBI excess being either the SZE or radio source contamination.

\end{abstract}


\section{Introduction}\label{sec:intro}

Observations of the cosmic microwave background (CMB) are among the most
powerful and important tests of cosmological theory. Measurements of the
angular power spectrum of CMB temperature anisotropies on angular scales $>
10^\prime$ 
- corresponding to multipoles $\ell \lesssim 1000$ - \citep{spergel06} in conjunction with other cosmological probes \citep{burles01,cole05,tegmark06,riess07} have produced compelling evidence for the $\Lambda$CDM cosmological model. At higher multipoles, measurements probe the Silk damping tail of the power spectrum and provide an independent check of the cosmological model.

At smaller angular scales, the primary CMB anisotropies originating
 at redshift z = 1100 are exponentially damped by photon diffusion.  This
 effect, known as Silk damping, makes secondary anisotropies - those induced
 along the line of sight at lower redshift - increasingly important at higher
$\ell$.  At 150$\,$GHz, for example, the Sunyaev-Zel'dovich effect (SZE) is expected
 to be brighter than the primary CMB anisotropy at $\ell \gtrsim 2500$.  The amplitude
 of the SZE depends sensitively on the amplitude of the matter perturbations,
 scaling as $\sigma_8^7$.  Measurements of the CMB power spectrum 
with sufficient sensitivity on arcminute scales 
not only extend tests of the $\Lambda$CDM
 model's ability to accurately predict the features in the power spectrum of
 primary CMB anisotropy, but also probe the epoch of cluster formation and
 provide an independent measure of $\sigma_8$.

 In this paper, we present the complete results of observations of CMB
 temperature anisotropies at 150 GHz with 5$^\prime$ resolution from the 
 Arcminute Cosmology Bolometer Array Receiver (ACBAR)
 experiment at the South Pole station.  Previous measurements of the CMB power  
 spectrum by ACBAR have been presented in \cite{kuo04} (hereafter K04)  
 and \cite{kuo07} (hereafter K07).  In addition, the angular power spectrum
 on these angular scales has been measured at 30 GHz by CBI   \citep{readhead04}, VSA \citep{dickinson04}, and BIMA \citep{dawson06}, and at 100
 and 150 GHz by QuAD \citep{quad07}.

To date, measurements at angular scales $<10^\prime$  have been
consistent with predictions of the primary anisotropy based on measurements
at larger angular scales, with one exception.  Both CBI~\citep{mason03,bond05} and BIMA~\citep{dawson06} observe excess power for $\ell > 2000$ at 30 GHz compared to 
the predictions of the $\Lambda$CDM  model.  
This excess can be explained by the SZE if $\sigma_8 \approx 1$, 
but this value is in tension with the best-fit WMAP5 value of  $\sigma_8 \approx 0.8$. 
In K07, we found that while the frequency dependence of the excess is consistent with the SZE, 
the ACBAR and CBI data could not be used to rule out radio source contamination or systematic errors
as the source of the CBI excess.  
Careful measurements over a broad range of frequencies and angular scales are needed to provide a definitive answer.

Current estimates of the primordial power spectrum are consistent with the predictions of slow-roll inflation for a nearly scale-invariant spectrum which may also include a small running of the spectral index. 
Sparked by the modest evidence for negative running in the WMAP first-year data, a number of authors have investigated how existing data sets limit the allowed inflationary scenarios \citep{peiris03,mukherjee03,bridle03,leach03}.
 Small-scale data extend the range over which the primordial power spectrum is
measured and can potentially yield information about the mechanism of inflation.

This is the third and final ACBAR power spectrum release.  The first
release in K04 analyzed two fields from the 2001 and 2002 seasons with
a conservative field differencing algorithm.  The second ACBAR power
spectrum, presented by K07, added two more fields from the 2002 season
and implemented an improved, undifferenced Lead-Main-Trail (no-LMT)
analysis of the dataset.  The results presented here improve on the
previous work in two ways.  First, we include seven additional fields
observed in the 2005 Austral winter.  These fields double the total
number of detector hours and substantially improve the precision of
the band-power estimates.  In particular, the new fields were selected
to dramatically expand ACBAR's sky coverage in order to reduce the
cosmic variance contribution to the uncertainty and to improve the multipole resolution on angular
scales below $\ell \lesssim 1800$.  This angular range covers the
third to fifth acoustic peaks, making it especially interesting for
constraining cosmological models.  Second, we implement a new
temperature calibration based on a comparison of CMB fluctuations as
measured by ACBAR and the WMAP satellite~\citep{hinshaw08}. This
improved calibration tightens constraints on cosmological models found
from the combination of high-$\ell$ ACBAR band-powers with low-$\ell$
results from other experiments.

This paper is organized as follows. In \S~\ref{sec:instrument} we
review the ACBAR instrument and the CMB observation program.
The analysis algorithm is explained 
in \S~\ref{sec:analysis}. Section \S~\ref{sec:calib} is an 
overview of the calibration;
the details of cross-calibration between 
WMAP5 and ACBAR are discussed in Appendix A. Information on ACBAR's beams can be found in \S~\ref{sec:beam}.
Systematic tests and foreground contamination are discussed in 
\S~\ref{sec:sys}. 
We present the band-power 
results in \S~\ref{sec:results}, including a discussion of the scientific
interpretation. 
The ACBAR band-powers are combined with the results of other experiments to place 
constraints on the parameters of cosmological models in 
\S~\ref{sec:parameters}.
In \S~\ref{sec:conclusion}, we summarize the main results of this paper.

\section{The Instrument And Observations}\label{sec:instrument}
The ACBAR receiver was designed to take advantage of the excellent observing conditions at the South Pole to make extremely deep maps of CMB anisotropies~\citep{runyan03a}. It observes from the Viper telescope, a 2.1m off-axis Gregorian with a beam size of $5^{\prime}$ at $150\,$GHz. 
The beams are swept across the sky at near-constant elevation by the motion of an actuated flat tertiary mirror. 

  The receiver contains 16 optically active bolometers cooled to 240 mK by a three-stage He$^3$-He$^3$-He$^4$ sorption refrigerator.  The results reported here are derived from the 150 GHz detectors: there were 4-150 GHz bolometers in 2001, 8 in 2002 and 2004, and 16 in 2005. The detectors were background limited at 150 GHz with a sensitivity of approximately 340 $\mu K\sqrt{s}$.

In total, ACBAR observed 10 independent CMB fields, detailed in Table~\ref{tab:fields}.  The power spectrum derived from portions of four fields, CMB2/4, CMB5, CMB6, and CMB7 was reported in K07. Since then, we have completed the analysis of six new fields observed in 2005 as well as additional observations of the original four fields.  Details of the instrument configuration and performance in the 2001 and 2002 seasons are given in \citet{runyan03a}, while additional details of the CMB observations, data reduction procedures, and beam maps can be found in K04 and K07.

\begin{deluxetable*}{lrccccc}
\tabletypesize{\scriptsize}
\tablewidth{420pt}
\tablecaption{CMB Fields } 
\tablehead{
\colhead{Field} & \colhead{RA (deg)} & \colhead{dec (deg)} & \colhead{Area (deg$^2$)} &\colhead{Year}&\colhead{\# of detectors}& \colhead{Detector hours} }

\startdata

CMB2(CMB4) & 73.963 & -46.268 & 26(17) &2001(2002)&4(8)& 2.0k(1.1k) \\
CMB5 & 43.371 & -54.836 & 28 &2002(2005)&8(16)& 13.2k(10.6k) \\
CMB6 & 32.693 & -50.983 & 23 &2002&8& 2.8k \\
CMB7(ext*) & 338.805 & -48.600 & 28(107)  &2002(2005)&8(16)& 3.4k(12.4k)\\
CMB8 & 82.297 & -46.598 & 61  &2005&16& 17.1k\\
CMB9* & 359.818 & -53.135 & 93  &2005&16& 4.0k\\
CMB10* & 19.544 & -53.171 & 93  &2005&16& 3.6k\\
CMB11* & 339.910 & -64.178 & 91  &2005&16& 4.9k\\
CMB12* & 21.849 & -64.197 & 92  &2005&16& 2.7k\\
CMB13* & 43.732 & -59.871 & 78  &2005&16& 7.6k\\
\enddata
\tablecomments{\small
The central coordinates and size of each CMB field observed by ACBAR.  The sixth column gives the number of $150\,$GHz detectors. The last column gives the detector integration time for each field after cuts. The detector sensitivity was comparable (within $\sim$10\%) between 2002 and 2005. The six largest fields (marked with a *) are used in the calibration to WMAP. Note that the 2005 observations extended the declination range of the CMB7 field, leading to the combined field CMB7ext. CMB2(CMB4) and CMB8 also partially overlap, but are analyzed separately for computational reasons.  Approximately $1/4$ of the CMB2(CMB4) scans have been discarded to eliminate the overlapping coverage. The listed numbers reflect this loss.}

\label{tab:fields}
\end{deluxetable*}


\begin{figure*}[ht!]
\resizebox{\hsize}{!}{
\includegraphics[angle=90]{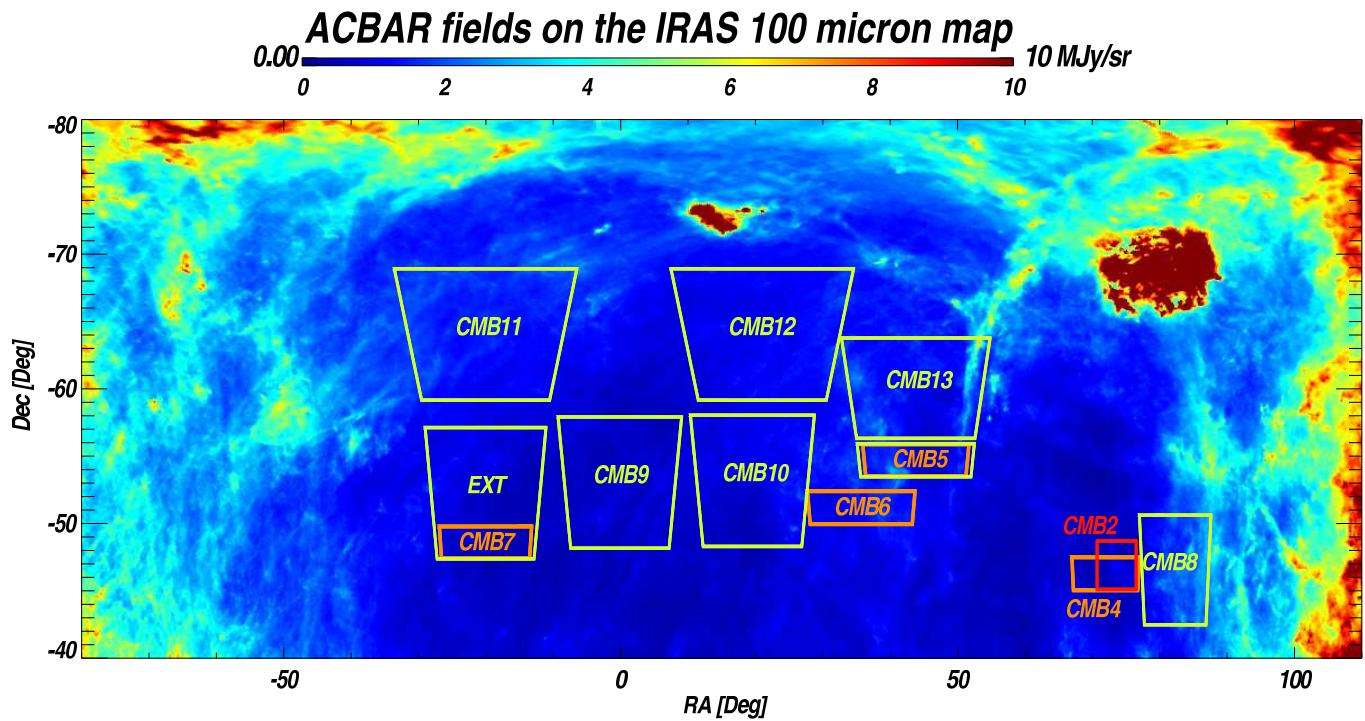}}

\caption{The ACBAR fields overlaid on the IRAS dust map.  The position of each field is plotted and labeled with the field name.
  The color coding indicates the year in which the observations occurred: red $\equiv$ 2001, orange $\equiv$ 2002, and yellow $\equiv$ 2005. The bulk of the 2005 season was targeted at large, comparatively shallow fields, increasing the total sky coverage by a factor of six.
The fields are plotted on top of the 100 $\mu$m IRAS dust map~\citep{schlegel98}.  Each field
lies within the ``Southern hole'', a region of low dust emission visible from the South Pole.  The CMB8 field (lower right corner) was targeted at the deep region of the B03 experiment as an alternative calibration path to the WMAP cross-calibration used for the results presented here.
}
\label{fig:acbarfields} 
\end{figure*}

\section{Un-differenced Power Spectrum Analysis}\label{sec:analysis}

Following the conventions of the previous data releases,
the band-powers ${\bf q}$ are reported in units of $\mu K^2$, 
and are used to parameterize the power spectrum according to
\begin{equation}
\ell(\ell+1)C_{\ell}/2\pi\equiv{\cal D}_{\ell}=\sum_Bq_B\chi_{B\ell}\;,\label{dl}
\end{equation}
where $\chi_{B\ell}$ are tophat functions; $\chi_{B\ell}=1$ for $\ell \in B$, and
$\chi_{B\ell}=0$ for $\ell \not\in B$.
The ACBAR observations were carried out in a {\em lead-main-trail}
(LMT) pattern. Originally, the three fields were differenced according to the formula $M-(L+T)/2$ in order to remove time-dependent 
chopper synchronous offsets. In K07, this conservative strategy was shown to be unnecessary and an un-differenced analysis algorithm was presented. 
We continued to observe in a lead-trail or LMT pattern in 2005 in order to produce maps wider than the maximum range ($\sim 3^\circ$) of the chopping tertiary mirror. The un-differenced analysis presented in K07, and used for this paper's analysis, is outlined below with any differences in the application to the 2005 data set highlighted.

Let $d_\alpha$ be a measurement of the CMB temperature at pixel $\alpha$. The data vector can be represented as the sum of the signal, noise and chopper synchronous offsets: $d_\alpha = s_\alpha + n_\alpha +o_\alpha$. For example, although the chopping mirror moves the beams at nearly-constant elevation, the slight residual atmospheric gradient produces a chopper synchronous signal $o_\alpha$ which is an approximately quadratic function of chopper angle.
To remove these offsets, the data from each chopper sweep
are filtered with the ``corrupted mode projection'' matrix
${\bf \Pi}$  to produce the cleaned
time stream ${\tilde{\bf d}}\equiv {\bf \Pi d}$.

The ${\bf \Pi}$ matrix projects out a third to tenth order polynomial which suppresses large angular scale chopper offsets. The order of the polynomial removed depends on the amplitude of atmosphere-induced cross-channel correlations. 
As described in K07, small angular scale offsets can be be removed by subtracting an ``average'' chopper function. 
For 2002 data, we removed a chopper synchronous offset from each data strip where the amplitude of the offset at each 
sample in the strip is free to vary quadratically with elevation in the map.  
The large fields observed in 2005 have up to four times the dec range of the fields observed in 2001 and 2002 ($\sim10^\circ$ vs. $\sim2.5^\circ$). 
For 2005 data, we allow the offset to vary from a third to fifth order polynomial depending on the extent of the map in declination.  
A zeroth order polynomial in elevation removes the average chopper function and the higher order terms effectively act as a high-pass filter on changes in the offset as a function of time or elevation.
This anisotropic filtering removes offset-corrupted modes while preserving most of the uncorrupted modes for the power spectrum analysis.
The loss of information at high-$\ell$ is small; the removed modes account for only a few percent of the total 
degrees of freedom of the data.

The corrupted mode projection matrix $\Pi$
can be represented as the product of two matrices, $\Pi\equiv\Pi_2 \Pi_1$.
The operator $\Pi_1$ is the original $\Pi$ matrix referenced in K04 which 
adaptively removes polynomial modes in RA.
The additional operator $\Pi_2$ removes modes in dec independently 
for each of the lead, main, and trail fields and can be further decomposed into the product $\Pi_2 = \Pi_2^{Poly} \Pi_2^{LPF}$. The operator $\Pi_2^{Poly}$ performs the aforementioned polynomial projection in dec to remove small-scale chopper offsets. The second operator $\Pi_2^{LPF}$ imposes a low-pass filter (LPF) $\ell<3200$ on each dec strip. The dec strips are perpendicular to the scan direction; the timestreams have always had a LPF applied in the scan direction.   The pixelation used when estimating the power spectrum is too large to resolve all of the noise power (at $\ell$ up to 10,800), causing out-of-band noise to be aliased into the signal band ($\ell<3000$) if a LPF is not applied.  Eliminating this high-frequency noise reduces the contribution of instrumental noise to the reported band-powers.

Using the pointing model, the cleaned timestreams are coadded  to create a map:
$$
  {\bf \Delta}={\bf L}{\bf d}.
$$
The noise covariance matrix of the map can be represented as
$$
{\bf C}_N={\bf L}\langle {\bf n n}^t \rangle {\bf L}^t.
$$ 
where ${\bf n}$ is the timestream noise.  The noise matrix is diagonalized as part of applying a high signal-to-noise transformation to the data. Eliminating modes with insignificant information content reduces the computational requirements of later steps in the analysis.

In order to apply the iterative quadratic band-power estimator, we need to know the partial derivative $\frac{\partial {\bf C}_T}{\partial q_B}$ of the theory covariance matrix ${\bf C}_T$ with respect to each band-power $q_B$.  The theory matrix can be calculated by considering the effects of the filtering on the raw sky signal.  The signal timestream $s_\alpha$ is the convolution of the true temperature map ${\mathfrak T}({\bf r})$ with the instrumental beam function $B_\alpha({\bf r})$
$$s_\alpha = \int d^2r {\mathfrak T}({\bf r}) B_\alpha({\bf r}).$$

The signal component of the coadded map will be ${\bf \Delta^{sig}}={\bf L}{\bf s}$ or 

$$
  \Delta_i^{sig}=\int d^2r F_i({\bf r}) {\mathfrak T}({\bf r}),\label{tsig}
$$
where we have defined the pixel-beam filter function $F_i$ 
$$
  F_i({\bf r})=\sum_\alpha L_{i\alpha}B_\alpha({\bf r}).
$$

The theory covariance matrix can be calculated in the flat sky case to be 
$$
C_{T\{ij\}} \equiv
\langle \Delta_i\Delta_j\rangle^{sig}=
\int\!\!\int d^2r d^2r' F_i({\bf r})F_j({\bf r}')
\langle {\mathfrak T}({\bf r}){\mathfrak T}({\bf r}')\rangle
$$
$$
=\int\!\!\int d^2r d^2r' F_i({\bf r})F_j({\bf r}') \int \frac{d^2\ell}{(2\pi)^2}
C_{\ell} \cdot e^{i{\bf l}\cdot ({\bf r}-{\bf r}')}
$$
\begin{equation}
=\int \frac{d^2\ell}{(2\pi)^2}C_{\ell} \cdot {\tilde F}_i^*({\bf l}){\tilde F}_j({\bf l}),
\label{ct} 
\end{equation}
where ${\tilde F}_i({\bf l})$ is the Fourier transform of the pixel-beam filter function. The partial derivative of the theory matrix can be calculated in a straightforward manner from equations \ref{dl} and \ref{ct}.

This algorithm does not require the instrument beams to remain constant.
The actual ACBAR beam sizes vary slightly with chopper angle \citep{runyan03a}.  The measured beam variations can be fit to a semi-analytic function as described in K04 to create a more accurate representation of the true beam shape across the map. We use the corrected beam sizes when removing point sources. In K04 and K07, we found that the differences in the power spectra from using the map-averaged beam or exact beam for each pixel were negligible. For the band-powers reported in Table \ref{tab:bands}, an averaged beam is used for the entire map.  

As in K07, we calculate the full two dimensional
noise correlation matrix directly from the time stream data without using 
Fourier transforms.
All the numerical calculations are performed on the 
National Energy Research Scientific Computing Center (NERSC) IBM SP RS/6000. 
The evaluation of the Fourier transform of $F_i({\bf r})$ is the most computationally expensive step of this analysis. We use an iterative quadratic estimator to find the maximum likelihood band-powers \citep{bond98}. 
The resulting band-powers are presented in Table \ref{tab:bands} and  Figure \ref{fig:acbar}.

\section{Calibration}\label{sec:calib}

We derive the absolute calibration of ACBAR by directly comparing the 2005 ACBAR maps to the WMAP5 V and W-band temperature maps \citep{hinshaw08}. We pass the WMAP5 maps through a simulated version of the ACBAR pipeline to ensure equivalent filtering and cross-spectra are calculated for each field.  The ratios of the cross-spectra are used to measure the relative calibration after being corrected for the respective instrumental beam functions. We had initially applied this calibration scheme to the WMAP3 maps.  Transitioning to the WMAP5 dataset lowered the calibration by 1.4\% in CMB temperature units and slightly reduced the overall uncertainty. The ACBAR band-powers are unchanged except for this calibration factor. Results for ACBAR's six largest fields (approximately $600\,{\rm deg}^2$ in area) are combined to achieve a calibration accuracy of 1.97\% for the 2005 data.

The 2005 calibration is transfered to 2001 and 2002 through a comparison of power spectra for overlapping regions observed by ACBAR in each year.  The CMB5 field is used to extend the calibration of the 2005 season to the 2002 data.  The CMB5 calibration is carried to other fields observed in 2002 by daily observations of the flux of RCW38. The calibration of the CMB4 field (observed in 2002) then is transfered to the 70\% overlapping CMB2 field (observed in 2001). According to the new calibration, results from the RCW38-based calibration used for the 2002 data in K07 need to be multipled by $0.959 \pm 0.032$. 
 Including the year-to-year calibration uncertainty, the final calibration has an uncertainty of 2.05\% in CMB temperature units (4.1\% in power).  Additional details of this procedure are discussed in Appendix \ref{app:calib}.

\section{Beam Determination}\label{sec:beam}

\begin{figure*}[htp]
\begin{center}
\plotone{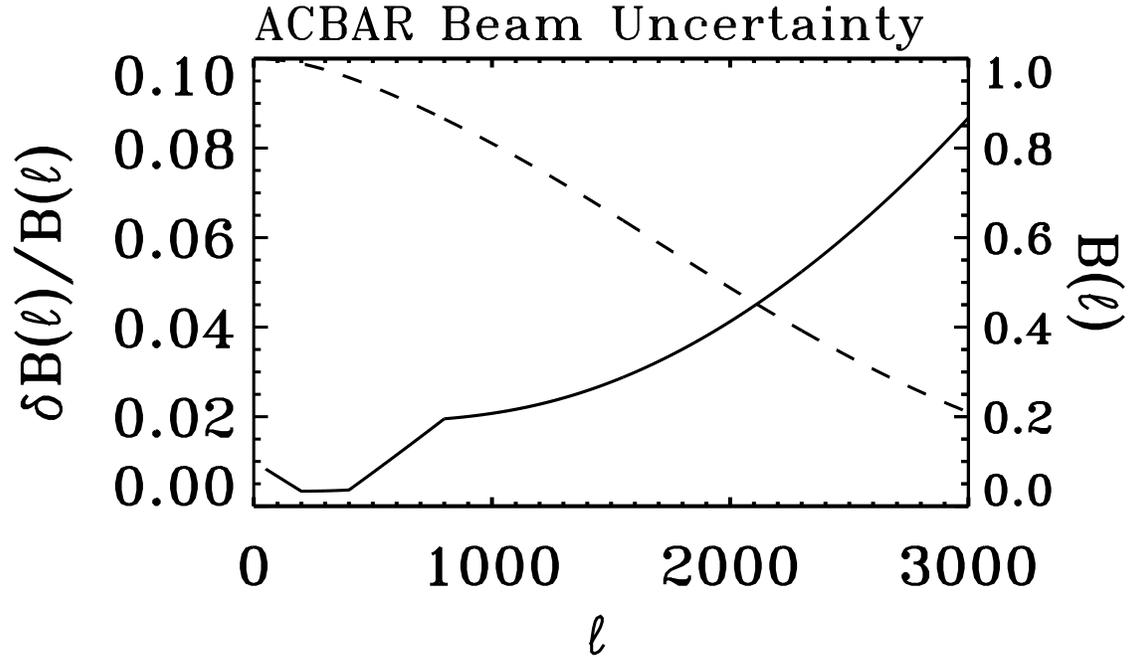}

\caption{ACBAR Beam Uncertainty and Beam Function. {\it Solid line \& left axis}:  The 1$\sigma$ envelope for uncertainty in the ACBAR beam function $B_\ell$.  The increasing uncertainty above $\ell=1000$ reflects the 2.6\% uncertainty in the fitted Gaussian FWHMs. The behavior at $\ell<1000$ is a combination of the uncertainty in the measured sidelobes and the calibration method `pinning' the transfer function for $\ell \in [256,512]$. {\it Dashed line \& right axis}: The measured ACBAR beam function.
 }\label{fig:blerr}
\end{center}
\end{figure*}

The beams are well-described by a symmetric Gaussian, with their main-lobe FWHM determined to  $2.6\%$ by continuous measurements of the images of bright quasars located in the CMB fields. The beam sidelobes were measured to the level of 30 dB with observations of Venus made in 2002.  Venus is extremely bright at millimeter wavelengths and with a diameter of $\lesssim 1^{\prime}$, is much smaller than ACBAR's beam size. However, there are extended periods during which ACBAR was unable to observe Venus. ACBAR observed RCW38, a bright HII region in the galactic plane, every day. We compare deep, coadded observations of RCW38 to constrain the temporal variability of the beam sidelobes when Venus was unavailable.  The complex structure surrounding RCW38 makes it difficult to directly recover the beam shape $B({\bf r})$. Instead, we monitor ratios of the beam-smoothed RCW38 maps $ \int d^2r S^{RCW38}({\bf r}) B({\bf r})$.  Any observed differences in the maps would indicate a change to the instrumental beam function as the morphology of RCW38's emission $S^{RCW38}$ is expected to be constant.  We set an upper limit on the possible temporal variations in the map and use this to constrain temporal variations in the beam function.  The estimated band-power uncertainty from the beam function is comparable to the overall calibration uncertainty and is plotted in Figure \ref{fig:blerr}.

\section{Systematic Uncertainties and Foregrounds}\label{sec:sys}

\subsection{Jackknife Tests}\label{subsec:jackknife}
We performed a series of tests to search for and constrain 
potential systematic errors in the power spectrum results. As described in K04, the data can be divided into two halves based on whether the chopping mirror is moving to the left or right.  The ``left minus right'' jackknife is a sensitive test for errors in the transfer function correction, microphonic vibrations excited by the chopper motion, or the effects of wind direction. Maps with bright sources such as RCW38 can provide particularly sensitive tests of the transfer function (see \cite{runyan03a} for a description of ACBAR's transfer functions).  
Similarly, the data can be split based on the time that the observation occurred.  A non-zero signal could be produced in the ``first half minus second half'' jackknife by variation in the calibration, pointing, beam and sidelobe, 
or any other time dependent variations in the instrument. In addition, the band-powers of each jackknife constrain 
the mis-estimation of noise during that period.

\begin{figure*}[htp]
\centering
\includegraphics[width=3.5in]{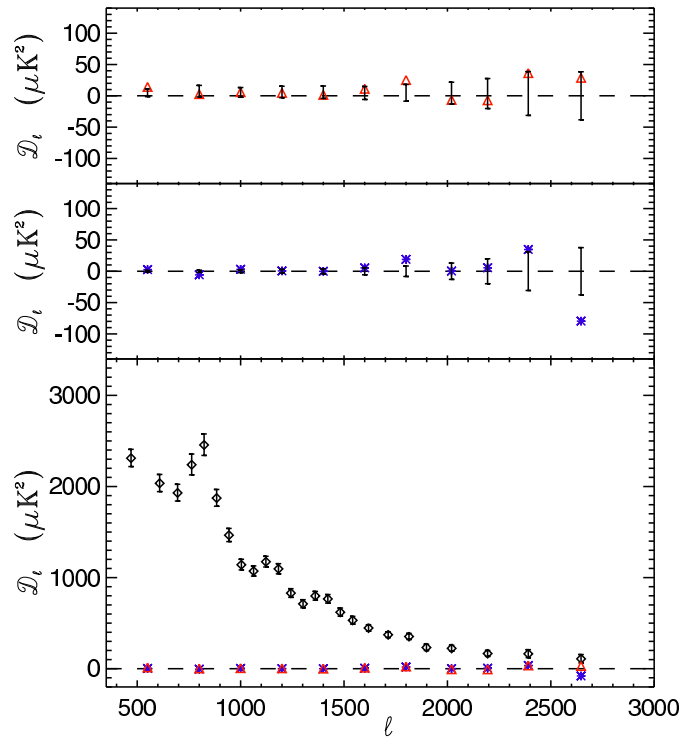}

\caption{Systematic tests performed on the ACBAR data. {\em Top}: Power spectrum ({\em red triangle}) for differenced maps from the first half of the season and second half of the season for each field, compared to the results of Monte Carlo simulations ({\em error bars}).
 {\em Middle}: Power spectrum ({\em blue star}) derived from difference maps of the left- and right-going chopper sweeps for all ten fields.
{\em Bottom}: The undifferenced band-powers from Table \ref{tab:bands} ({\em black diamonds}) compared to both jackknife power spectra: the  left-right jackknife ({\em blue star}) and first half-second half jackknife ({\em red triangle}). 
 }\label{fig:sys}
\end{figure*}

We applied the left-right jackknife to the 2005 CMB data and
found the band-powers were inconsistent with zero at 2.5$\sigma$ at high-$\ell$ ($\ell > 2100$).  We reran a set of left-right jacknives dropping individual channels, and found that two channels stood out.  With both channels excluded, the discrepancy in the left-right jackknife band-powers disappeared. We were unable to find evidence for unusual microphonic lines or transfer functions in the two problematic channels, but hypothesize that these two channels have subtle microphonic response in the signal bandwidth that are detectable only in a deep integration. Both channels are excluded from the 2005 data for all results reported in this paper.

We apply the first-second half jackknife test to the joint CMB power spectrum with the exclusion of the bad channels from the 2005 data.
The power spectrum of the chronologically differenced maps is compared to the band-powers of a set of Monte Carlo realizations of simulated difference maps in order to account for a number of effects such as the small filtering differences due to different scan patterns and the temporal uncertainty in the beam sidelobes (see \S\ref{sec:beam}).  We find that the jackknife band-powers are consistent with the predictions of the Monte Carlo above $\ell = 400$.  There is a 4$\sigma$ residual of $\sim$15 $\mu$K$^2$ in the first bin.  Because the combined statistical and cosmic variance uncertainty in this bin is a factor of six larger, we assume that the band-power estimate will not be significantly biased.

We also perform the left-right jackknife on the joint CMB power spectrum found from the complete data set. The results are consistent with zero for $\ell > 900$. Statistically, the probability to exceed the measured $\chi^2$ for $\ell > 900$ is 15\%. The results are inconsistent with zero at a very low ($\sim$4 $\mu K^2$) level (Fig. \ref{fig:sys}) on larger angular scales. This residual could be due to a small noise mis-estimate at low-$\ell$, possibly caused by neglected atmospheric correlations.   The jackknife failure of $\sim$4 $\mu K^2$ is much smaller than the band-power uncertainties (90 - 300 $\mu K^2$) in these $\ell$-bins which are dominated by cosmic variance. The first-second half jackknife is insensitive to discrepancies of this magnitude due to the greater uncertainties introduced by small pointing and filtering 
differences.  
We conclude that the complete ACBAR data set shows no significant residuals in the jackknife tests and we 
expect no significant systematic contamination of the resulting power spectrum.

\subsection{Foregrounds}\label{subsec:foregrounds}

The potential contribution of foreground emission must be considered in the interpretation of CMB temperature anisotropies.  There are three foregrounds with significant emission at $150\,$GHz on the relevant angular scales: galactic dust, extragalactic radio sources, and dusty proto-galaxies. As an effectively single-frequency instrument, ACBAR depends on data from other experiments to construct and constrain foreground models.  We use the methodology described in K04 to remove templates for radio sources and dust emission from the CMB maps without making assumptions about their flux.  
The contribution from dusty protogalaxies is less certain; however, we do not expect the combined residual foreground 
emission to significantly impact the power spectrum for $\ell < 2400$.

We remove modes from the CMB maps corresponding to radio sources in the $4.85\,$GHz Parkes-MIT-NRAO (PMN) survey \citep{wright94}. 
Of the 1601 PMN sources with a flux greater than $40\,$ mJy that lie in the ACBAR fields, we detect 37 sources including the guiding quasars at greater than 3$\sigma$ with the application of an optimal matched filter.  There are less than 2.2 false detections expected with this detection threshold. The measurement errors are estimated through sampling the distribution of pixels in a set of 100 Monte Carlo realizations of the CMB+noise for each field. Table \ref{tab:pmnsources} lists the parameters of the detected PMN sources. 
Except for the few bright sources detected at $150\,$GHz, removing the PMN point sources does not 
significantly affect the band-powers.

It is possible that faint radio sources, undetected at $150\,$GHz, could contribute to the observed band-powers. We parameterize the contribution as 
\begin{equation}\label{eq:src}
{\cal D}^{\rm src}_\ell = q_{\rm src}\left(\frac{\ell}{2600}\right)^2 \, \mu K^2\, 
\end{equation}
which is appropriate for unclustered point sources.
We compare the $150\,$GHz ACBAR point source number counts to the model in \citet{white04} based on WMAP Q-band data, $$\frac{dN}{dS_\nu} = \frac{80~deg^{-2}}{1~mJy} \left(\frac{S_\nu}{1~mJy}\right)^{-2.3},$$ to constrain the residual power contribution at $150\,$GHz.
We can use this model to estimate the residual band-power contribution due to sources too faint to be included 
in the PMN catalog.

Following the convention in that work, the spectral dependence of the fluxes is parameterized as $S_\nu \propto \nu^\beta$. 
The number of PMN sources detected at $150\,$GHz in a logarithmic flux bin, $n^{obs}_B$, are  compared to the predicted number counts from the same population of sources with a given $\beta$, $n^{(\beta)}_B + n^{noise}_B$.  Here, $n^{(\beta)}_B$ is the modeled number counts and $n^{noise}_B$ is the expected number of false detections due to ACBAR's measurement uncertainty.  The number counts are assumed to follow a Poisson distribution. Sources with estimated measurement errors greater than 140 mJy  in the ACBAR maps are cut to reduce the $n^{noise}_B$ term. The modeled number counts $n^{(\beta)}_B$ are scaled by $(N_{tot}-N_{cut})/N_{tot}$ to compensate. All other sources with measured amplitudes greater than 350 mJy at $150\,$GHz are included in the calculation without consideration of the signal-to-noise.  
If the sources found by ACBAR at $150\,$GHz are the same population found by WMAP, this implies a uniform spectral index of $\beta = 0.14\pm 0.15$. However, this small sample selected for high flux at $150\,$GHz is heavily biased toward sources with flat or rising spectra.  We increase ACBAR's sensitivity to dimmer sources by binning all sources within a given PMN flux range, and look at the ratio of the average flux at $150\,$GHz to the average flux at 4.85 GHz within each bin. We find the ratio ($S_{150}/S_{4.85}$) increases with PMN flux from 0.07 for sources below 400 mJy  to 0.41 for sources above 1600~mJy at 4.85~GHz.  
This implies that the sources in the PMN catalog have a flux-dependent spectral index where dimmer objects typically have a more steeply falling spectrum. 
The band-power contribution of the low-flux sources depends sensitively on the extrapolation of the 40~mJy flux cutoff in the $4.85\,$GHz PMN catalog to $150\,$GHz.  Based on the observed flux ratios for PMN sources with $S_{4.85} < 400$ mJy, we conservatively assume  $S_{150}/S_{4.85} = 0.1$ for a flux cutoff at $4\,$mJy at $150\,$GHz.   This flux ratio corresponds to a spectral index of $\beta = -0.67$, well below $\beta = 0.14\pm 0.15$ estimated from the ACBAR detected sources and the WMAP Q-band source model. 
Estimating the residual radio source band-power contribution at $150\,$GHz with a flux cutoff of $4\,$mJy gives $q_{\rm src}^{radio} = 2.2$.  At this level, the residual contribution from radio sources will be negligible in the ACBAR data.

\begin{figure*}[ht!]
\epsscale{1.}
\plotone{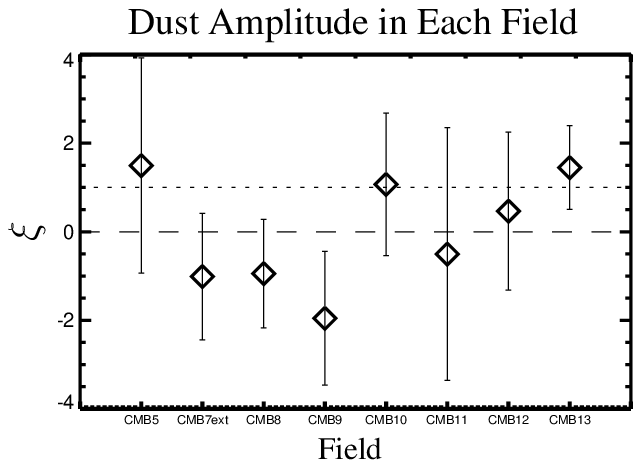}
\caption{Dust emission is not detected in the ACBAR fields. Parameterizing the dust signal as $T_{CMB}+\xi T_{FDS}$, a suite of Monte Carlo realizations of maps of CMB and noise is used to estimate the uncertainty in $\xi$. We find the upper limits in each field to be consistent with the FDS99 model ($\xi = 1$), but the data somewhat favor a lower dust amplitude. The reduced $\chi^2$ of the measured amplitudes $\xi$s is 0.75 under the assumption that $\langle \xi \rangle = 0$ (the dashed line).  The reduced $\chi^2$ for the FDS99 model with $\langle \xi \rangle = 1$  is 1.12 (the dotted line).
}
\label{fig:dustamp} 
\end{figure*}

The ACBAR fields are positioned in the ``Southern Hole,'' a region of exceptionally low Galactic dust emission (Figure~\ref{fig:acbarfields}).  \citet{finkbeiner99} (FDS99) constructed a multi-component dust model that predicts thermal emission at CMB frequencies from the combined observations of IRAS, COBE/DIRBE, and COBE/FIRAS.  Taking into account the ACBAR filtering, the FDS99 model\footnote{We use the default model 8 of FDS99.} predicts a RMS dust signal at the few $\mu K$ level primarily on large angular scales. The ACBAR maps can be decomposed as the sum of the CMB and dust signals $T_{CMB}+\xi T_{FDS}$.   The dust amplitude parameter $\xi$ is predicted to equal unity by the FSD99 model. The ACBAR maps are cross-correlated with the dust templates $T_{FDS}$ to calculate the amplitude in each field.   The errors are estimated by applying the same procedure to 100 CMB+noise map realizations for each field.  The uncertainty in $\xi$  is dominated by CMB fluctuations. The best-fit amplitude from combining all the fields is $\xi = 0.1 \pm 0.5$. The estimated amplitudes of the individual fields are shown in Figure \ref{fig:dustamp}.  The reduced $\chi^2$ of the measured amplitudes $\xi$s  of the eight fields analyzed is 0.75 for the no-dust assumption of $\langle \xi \rangle = 0$ and increases to $\chi^2 = 1.12$ for the FDS99 model amplitude of  $\langle \xi \rangle = 1$. Therefore, the ACBAR data slightly favor a lower amplitude than predicted by the FDS99 model. The dust signal is not detectable in any of the ACBAR fields, and removing the dust template has a negligible impact on the measured power spectrum.  

Dusty IR galaxies are the third and least constrained foreground in the ACBAR fields.  This population of high-redshift, star-forming galaxies has been studied by several experiments at higher frequencies \citep{coppin06,laurent05,maloney05,greve04}.  However, as discussed in K07, extrapolating the expected signal to $150\,$GHz remains highly uncertain, and there remain significant uncertainties in the number counts $\frac{dN}{dS}$ and spatial clustering of the sources. The frequency dependence can be empirically determined by comparing the measured number counts in overlapping fields observed at different frequencies.  This comparison has been done with MAMBO (1.2 mm) and SCUBA (850 $\mu$m), leading to a spectral dependence of $S_\nu\propto \nu^{2.65}$ \citep{greve04}.  A second method of estimating the index used nearby galaxy data to obtain $S_\nu \sim \nu^{2.6\pm 0.3}$ \citep{knox04}. 
The uncertainty in the spectral dependence significantly affects the extrapolation of the flux of dusty galaxies to $150\,$GHz. We use estimates of the source number counts from the SHADES survey~\citep{coppin06} and Bolocam Lockman Hole Survery~\citep{maloney05}. We apply the formulas in \citet{scott99} to estimate the expected power spectrum for the source number counts, ignoring the clustering terms. In this limit, ${\cal D}_{\ell}$ will have the form in eq. \ref{eq:src}. Scaling the results to $150\,$GHz with the MAMBO/SCUBA prescription of $S_\nu\propto \nu^{2.65}$ leads to an estimated contribution of  $q_{\rm src}^{dusty} = 17-29$.  This range reflects the differences between the measured number counts, but does not include the uncertainty in the spectral dependence of the fluxes.
Combining the median index with earlier SCUBA data fit by two power laws in $S$ from
\citep{borys03}, we find an excess of 22  $\mu {\rm K}^2$ at $\ell = 2600$, within that range.
This level is only a factor of two smaller than the instrumental noise of ACBAR and might
influence the interpretation of high-$\ell$ band-powers.  
We tentatively assume that contamination from dusty proto-galaxies does not significantly effect the 
resulting power spectrum.
The implications of relaxing this assumption for cosmological parameter estimation are 
explored in \S~\ref{subsec:ressourcemarg}.

 
\begin{deluxetable*}{llrrr}
\tabletypesize{\small}
\tablewidth{360pt}
\tablecaption{Millimeter Bright PMN Sources }
\tablehead{
\colhead{Source Name/Position} & \colhead{Field} & \colhead{$S_{4.85}$ (mJy)}& \colhead{$S_{150}$ (mJy)}&\colhead{$\alpha_{150/4.85}$}}

\startdata
 PMN J0455-4616$^{*\circ}$ &       CMB2 &   1653 & $  2898 \pm  60 $ &  0.16 \\ 
 PMN J0439-4522 &       CMB2 &    634 & $   383 \pm  73 $ & -0.15 \\ 
 PMN J0451-4653 &       CMB2 &    541 & $   360 \pm  58 $ & -0.12 \\ 
 PMN J0253-5441$^{*\circ}$  &       CMB5 &   1193 & $  1260 \pm  62 $ &  0.02 \\ 
 PMN J0223-5347 &       CMB5 &    397 & $   176 \pm  27 $ & -0.24 \\ 
 PMN J0229-5403 &       CMB5 &    242 & $   147 \pm  17 $ & -0.14 \\ 
 PMN J0210-5101$^{*\circ}$  &       CMB6 &   3198 & $  1268 \pm  86 $ & -0.27 \\ 
 PMN J2207-5346$^{*\circ}$  &    CMB7ext &   1410 & $   381 \pm  68 $ & -0.38 \\ 
 PMN J2235-4835$^{*\circ}$  &    CMB7ext &   1104 & $  1509 \pm  75 $ &  0.09 \\ 
 PMN J2239-5701$^{*\circ}$  &    CMB7ext &   1063 & $   501 \pm  68 $ & -0.22 \\ 
 PMN J2246-5607 &    CMB7ext &    618 & $   386 \pm  51 $ & -0.14 \\ 
 PMN J2309-5703 &    CMB7ext &     56 & $   257 \pm  76 $ &  0.44 \\ 
 PMN J0519-4546$^{*\circ}$  &       CMB8 &  15827 & $  1375 \pm 101 $ & -0.71 \\ 
 PMN J0519-4546$^{*\circ}$  &       CMB8 &  14551 & $  1148 \pm  86 $ & -0.74 \\ 
 PMN J0538-4405$^{*\circ}$  &       CMB8 &   4805 & $  7114 \pm  87 $ &  0.11 \\ 
 PMN J0515-4556$^{*\circ}$  &       CMB8 &    990 & $   671 \pm  96 $ & -0.11 \\ 
 PMN J0526-4830 &       CMB8 &    425 & $    82 \pm  25 $ & -0.48 \\ 
 PMN J0525-4318 &       CMB8 &    217 & $    99 \pm  25 $ & -0.23 \\ 
 PMN J0531-4827 &       CMB8 &    142 & $    96 \pm  25 $ & -0.11 \\ 
 PMN J2357-5311$^{*\circ}$  &       CMB9 &   1782 & $   347 \pm  49 $ & -0.48 \\ 
 PMN J2336-5236 &       CMB9 &   1588 & $   233 \pm  57 $ & -0.56 \\ 
 PMN J2334-5251 &       CMB9 &    557 & $   432 \pm  57 $ & -0.07 \\ 
 PMN J0018-4929 &       CMB9 &    142 & $   178 \pm  57 $ &  0.07 \\ 
 PMN J0026-5244 &       CMB9 &     40 & $   192 \pm  62 $ &  0.46 \\ 
 PMN J0050-5738$^{*\circ}$  &      CMB10 &   1338 & $   773 \pm 108 $ & -0.16 \\ 
 PMN J0058-5659$^{*\circ}$  &      CMB10 &    739 & $   514 \pm  61 $ & -0.11 \\ 
 PMN J0133-5159$^{*\circ}$  &      CMB10 &    672 & $   248 \pm  72 $ & -0.29 \\ 
 PMN J0124-5113$^{*\circ}$  &      CMB10 &    308 & $   335 \pm  48 $ &  0.02 \\ 
 PMN J2208-6404 &      CMB11 &     53 & $   136 \pm  44 $ &  0.27 \\ 
 PMN J0103-6438 &      CMB12 &    395 & $   268 \pm  65 $ & -0.11 \\ 
 PMN J0144-6421 &      CMB12 &    152 & $   184 \pm  59 $ &  0.06 \\ 
 PMN J0303-6211$^{*\circ}$  &      CMB13 &   1862 & $   423 \pm  63 $ & -0.43 \\ 
 PMN J0309-6058$^{*\circ}$  &      CMB13 &   1103 & $   596 \pm  80 $ & -0.18 \\ 
 PMN J0251-6000 &      CMB13 &    433 & $   189 \pm  34 $ & -0.24 \\ 
 PMN J0236-6136 &      CMB13 &    406 & $   365 \pm  33 $ & -0.03 \\ 
 PMN J0257-6112 &      CMB13 &    178 & $   104 \pm  33 $ & -0.16 \\ 
 PMN J0231-6036 &      CMB13 &    174 & $   105 \pm  33 $ & -0.15 \\ 

\enddata
\tablecomments{\small
These sources from the PMN $4.85\,$GHz catalog are detected at 
$>3.0\sigma$ significance with ACBAR, corresponding to a false detection rate of $< 2.2$.  The fluxes at $4.85\,$GHz ($S_{4.85}$, from \citet{wright94}) and $150\,$GHz ($S_{150}$, measured by ACBAR) are given. For ACBAR, the flux conversion factor is $1~\mu K_{CMB} = 0.9$ mJy. 
The spectral index $\alpha$ is defined as $S_\nu\propto \nu^{\alpha}$. 
The flux of some of these sources varied by up to 50\% between years; this variability is not reflected in the estimated errors. The central guiding quasars (one in each of the 5 deeper fields) are marked with asterisks ($^*$).
These sources, as well as all other PMN sources $>40\,$mJy, are projected out from the data using the methods described by K04 and do not contribute to the 
power spectrum results. The brightest sources are marked with circles ($^\circ$) and are removed from the maps in a beam-independent method. Note that PMN J0519-4546a/b are within one beam width of each other and are not separately resolved by ACBAR.  As a result, the listed $\alpha$ for PMN J0519-4546a/b  is estimated from the sum of the fluxes at 4.85 GHz and the mean of the fluxes at $150\,$GHz.}
\label{tab:pmnsources}
\end{deluxetable*}

\section{Results And Discussions}\label{sec:results}

\subsection{Power Spectrum}\label{sec:ps}

\begin{figure*}[ht!]
\resizebox{\hsize}{!}{
\plotone{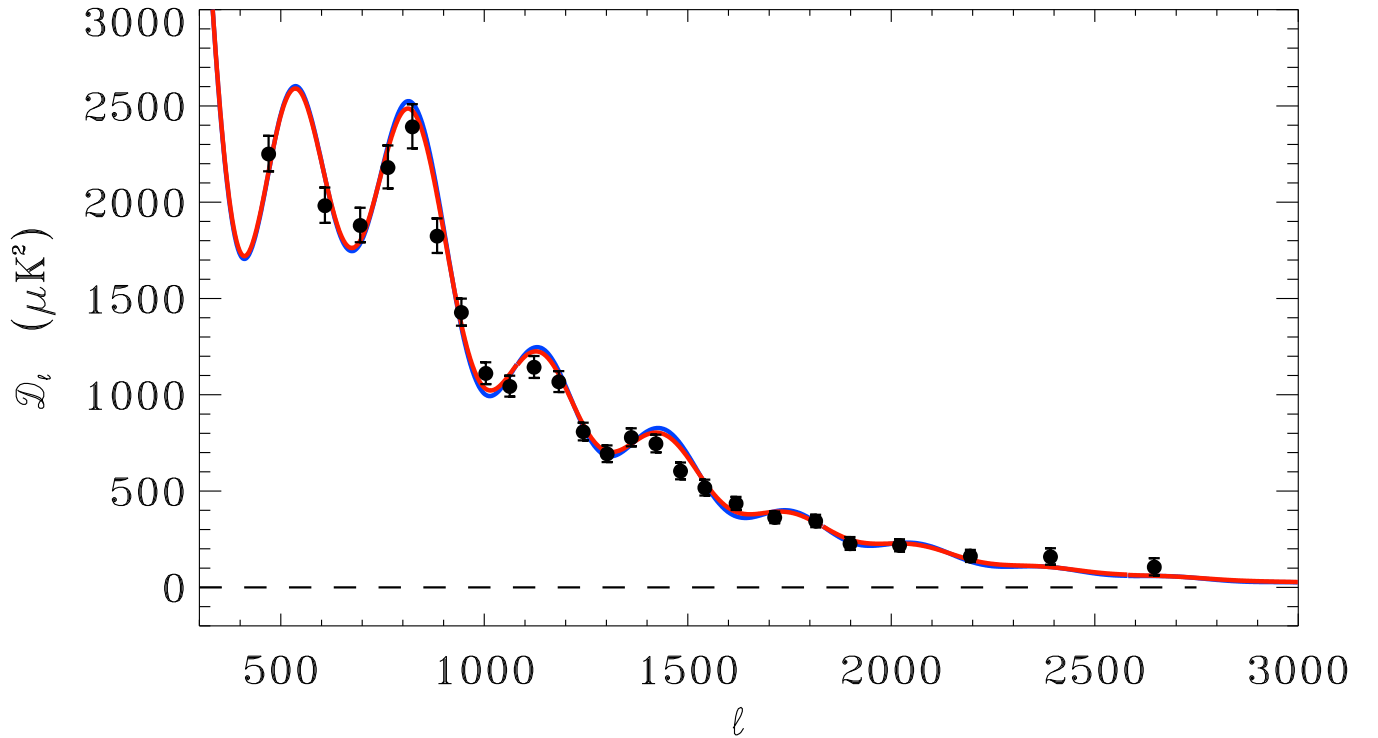}}
\caption{The decorrelated ACBAR band-powers for the full data set. The
  $1\sigma$ error bars are derived from the offset-lognormal fits to
  the likelihood function. The band-powers are in excellent agreement
  with a $\Lambda$CDM model.  The damping of the anisotropies is
  clearly seen with a S/N $>$ 4 out to $\ell=2500$.  The third
  acoustic peak (at $\ell \sim 800$), fourth acoustic peak (at $\ell
  \sim 1100$), and fifth acoustic peak (at $\ell \sim 1400$) are
  visible. The plotted lines are the best fits to the ACBAR and WMAP5
  band-powers for a spatially flat, $\Lambda$CDM universe with no SZE
  contribution. A lensed ({\it red}) and unlensed ({\it blue}) model
  spectrum is shown for a fixed parameter set; the lensed spectrum is a significantly better fit
  to the ACBAR data.  }
\label{fig:acbar} 
\end{figure*}

\begin{figure*}[ht!]
\resizebox{\hsize}{!}{
\plotone{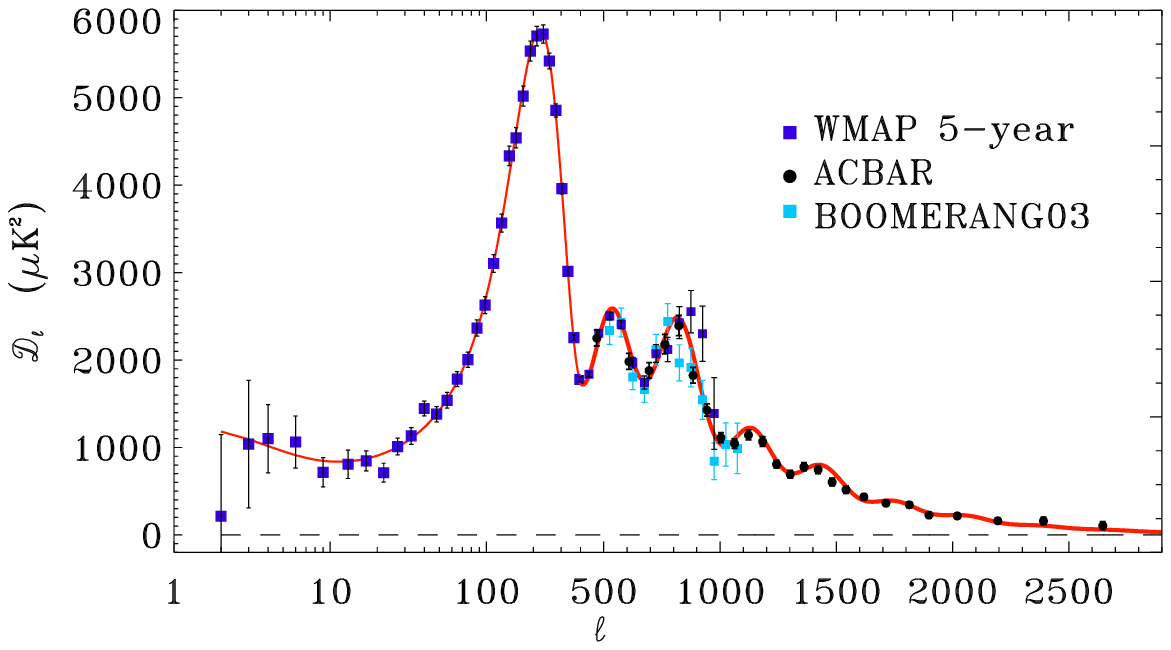}} 
\caption{\protect\small
The ACBAR band-powers plotted with those from WMAP5 \citep{hinshaw08} 
and the 2003 flight of BOOMERANG \citep{jones06}. The three 
experiments show excellent agreement in the region where they overlap.}
\label{fig:acbar_ext}
\end{figure*}

The power spectrum presented in Figure~\ref{fig:acbar} is produced by
the application of the analysis algorithm outlined in
\S~\ref{sec:analysis} to ACBAR data from the 2001, 2002 and 2005
austral winters. The resulting power spectrum is compared to the WMAP5
and B03 spectra in Figure~\ref{fig:acbar_ext}. The zero-curvature,
$\Lambda$CDM ``ACBAR+WMAP5" best fit model is shown in each figure for
reference. The decorrelated band-powers are tabulated in
Table~\ref{tab:bands}.  Our choice of the decorrelation
transformations follows \citet{tegmark97b}.  The band-powers can be
compared to a theoretical model using the window functions
\citep{knox99}.  As in K04, we sample the likelihood function ${\cal
  L}(\Delta) = \frac{1}{\sqrt{C}}e^{-(\Delta^t C^{-1} \Delta)/2}$ near
the maximum and fit the results with offset lognormal functions
\citep{bond2000}. The fit parameters ${\bf \sigma}, {\bf x}$ are
listed in Table~\ref{tab:bands} as well. The band-powers, likelihood
fit parameters, and window functions are available for download from
the ACBAR
website\footnote{http://cosmology.berkeley.edu/group/swlh/acbar/index.html}.

The ACBAR data extend the measurement of the temperature anisotropies
well into the damping tail with S/N $>$ 5 for $\ell \lesssim 2300$.
The fourth and fifth acoustic peaks are detected for the first time in
the ACBAR band-powers, providing additional support for the coherent
origin of anisotropy \citep{albrecht96}.  The position of the third
acoustic peak is consistent with previous detections of the feature by
CBI \citep{readhead04}, B03 \citep{jones06}, ACBAR (K07), and QUaD
\citep{quad07}. The ACBAR band-powers are in excellent agreement with
the cosmological models constrained by observations on larger angular scales. The
probability to exceed the reduced $\chi^2$ between the ACBAR
band-powers and the WMAP3 only best-fit $\Lambda$CDM model is
17\%. This probability increases to 62\% with the WMAP5 only best-fit model.
This serves as both a powerful confirmation of our basic cosmological model and an 
indication of the quality of the ACBAR data set.

\begin{deluxetable*}{ccccc}
\tabletypesize{\small}

\tablewidth{270pt}

\tablecaption{Joint Likelihood Band-powers }

\tablehead{\colhead{$\ell$ range}&\colhead{$\ell_{eff}$} &\colhead{$q$ ($\mu{\rm K}^2$)}& \colhead{$\sigma$ ($\mu{\rm K}^2$)} &
\colhead{x ($\mu{\rm K}^2$)}}
\startdata
350-550 & 470 & 2250 & 92 & -345 \\
550-650 & 608 & 1982 & 92 & -303 \\
650-730 & 694 & 1879 & 89 & -267 \\
730-790 & 763 & 2180 & 111 & -239 \\
790-850 & 823 & 2391 & 115 & -255 \\
850-910 & 884 & 1824 & 90 & -164 \\
910-970 & 944 & 1427 & 70 & -104 \\
970-1030 & 1003 & 1111 & 57 & -18 \\
1030-1090 & 1062 & 1043 & 54 & 17 \\
1090-1150 & 1122 & 1143 & 57 & 34 \\
1150-1210 & 1182 & 1067 & 54 & 75 \\
1210-1270 & 1242 & 808 & 46 & 119 \\
1270-1330 & 1301 & 693 & 43 & 154 \\
1330-1390 & 1361 & 778 & 47 & 193 \\
1390-1450 & 1421 & 746 & 46 & 218 \\
1450-1510 & 1481 & 604 & 44 & 241 \\
1510-1570 & 1541 & 517 & 41 & 229 \\
1570-1650 & 1618 & 435 & 34 & 261 \\
1650-1750 & 1713 & 363 & 30 & 242 \\
1750-1850 & 1814 & 344 & 32 & 264 \\
1850-1950 & 1898 & 227 & 33 & 170 \\
1950-2100 & 2020 & 217 & 31 & 203 \\
2100-2300 & 2194 & 162 & 31 & 244 \\
2300-2500 & 2391 & 159 & 43 & 357 \\
2500-3000 & 2646 & 105 & 45 & 560 \\
\enddata
\tablecomments{\small
Band multipole range and weighted value $\ell_{eff}$, decorrelated band-powers $q_B$, 
uncertainty $\sigma_B$
, and log-normal offset $x_B$ from the
joint likelihood analysis of the 10 ACBAR fields. The positive trend in the log-normal offsets with increasing $\ell$ is due to the increasing contribution of instrumental noise to the error budget; the log-normal offset approaches infinity in the Gaussian limit.
The PMN radio point source and IRAS dust foreground templates have been
projected out in this analysis.}
\label{tab:bands}
\end{deluxetable*}


\subsection{Anisotropies at $\ell>2000$} 
\label{sec:excess}

Several theoretical calculations \citep{cooray00,komatsu02} and
hydrodynamical simulations \citep{bond02,white02} suggest that the
thermal Sunyaev-Zel'dovich effect power spectrum will exceed that of
the primary CMB temperature anisotropies for $\ell \gtrsim 2500$ at
150 GHz.  The amplitude of the SZE power spectrum is closely related to
the amplitude of matter perturbations which is commonly parameterized
as $\sigma_8$; the SZE power spectrum is expected to scale as
$\sigma_8^7$ \citep{zhang02}.  To a lesser extent, the level of the SZE
will also depend on details of cluster gas physics and thermal
history. The non-relativistic thermal SZE ($\Delta T_{SZ}$) has a
unique frequency dependence
\begin{equation}
\frac{\Delta T_{SZ}}{ T_{CMB}} = 
y \left(x\frac{e^x + 1}{e^x - 1} - 4\right),\,\label{szs}
 \end{equation}
 where $x = \frac{h\nu}{ k T_{\rm CMB}}$ = $\nu/56.8\,$GHz. The
 variable y is the Compton parameter and is proportional to the
 integrated electron pressure along the line of sight. The CBI
 extended mosaic observations \citep{readhead04} detected more power
 above $\ell = 2000$ than is expected from primary CMB anisotropies.
 This excess could be the first detection of the SZE power spectrum
 \citep{mason03,readhead04,bond02}.  However, there are alternative
 explanations for the observed power ranging from an unresolved
 population of low-flux radio sources to non-standard inflationary
 models \citep{cooray02,voids,Bfields} that produce higher than
 expected CMB anisotropy power at small angular scales.  The frequency
 dependence of the excess power can be exploited to help discriminate
 between the SZE and other potential explanations.

\begin{figure*}[ht!]
\resizebox{\hsize}{!}{
\plottwo{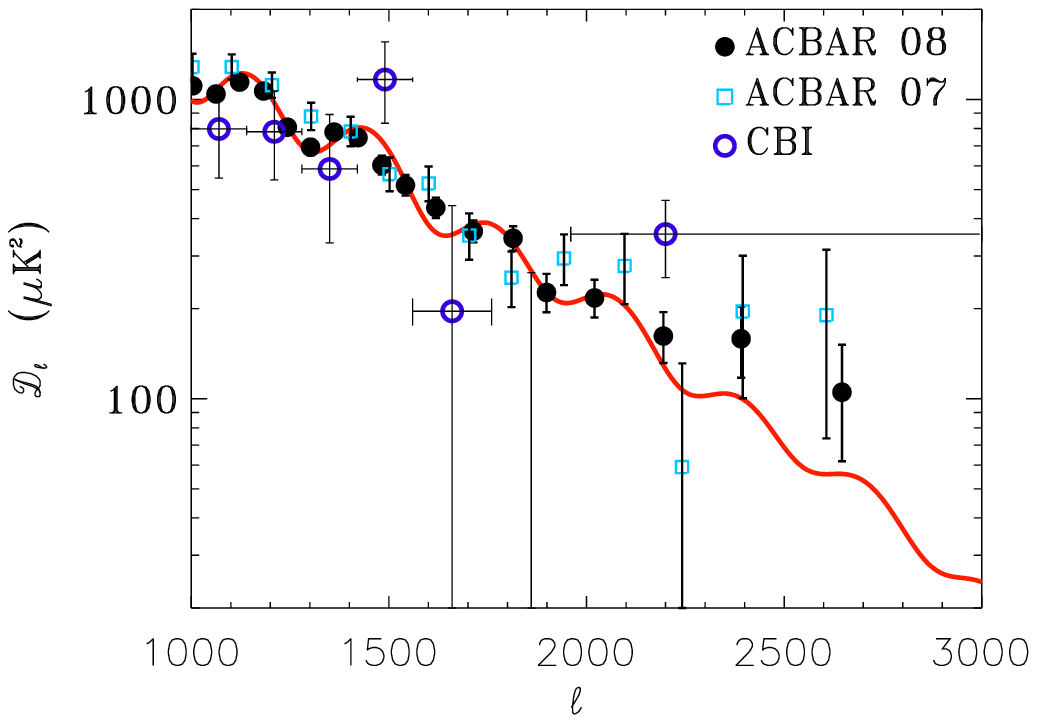}{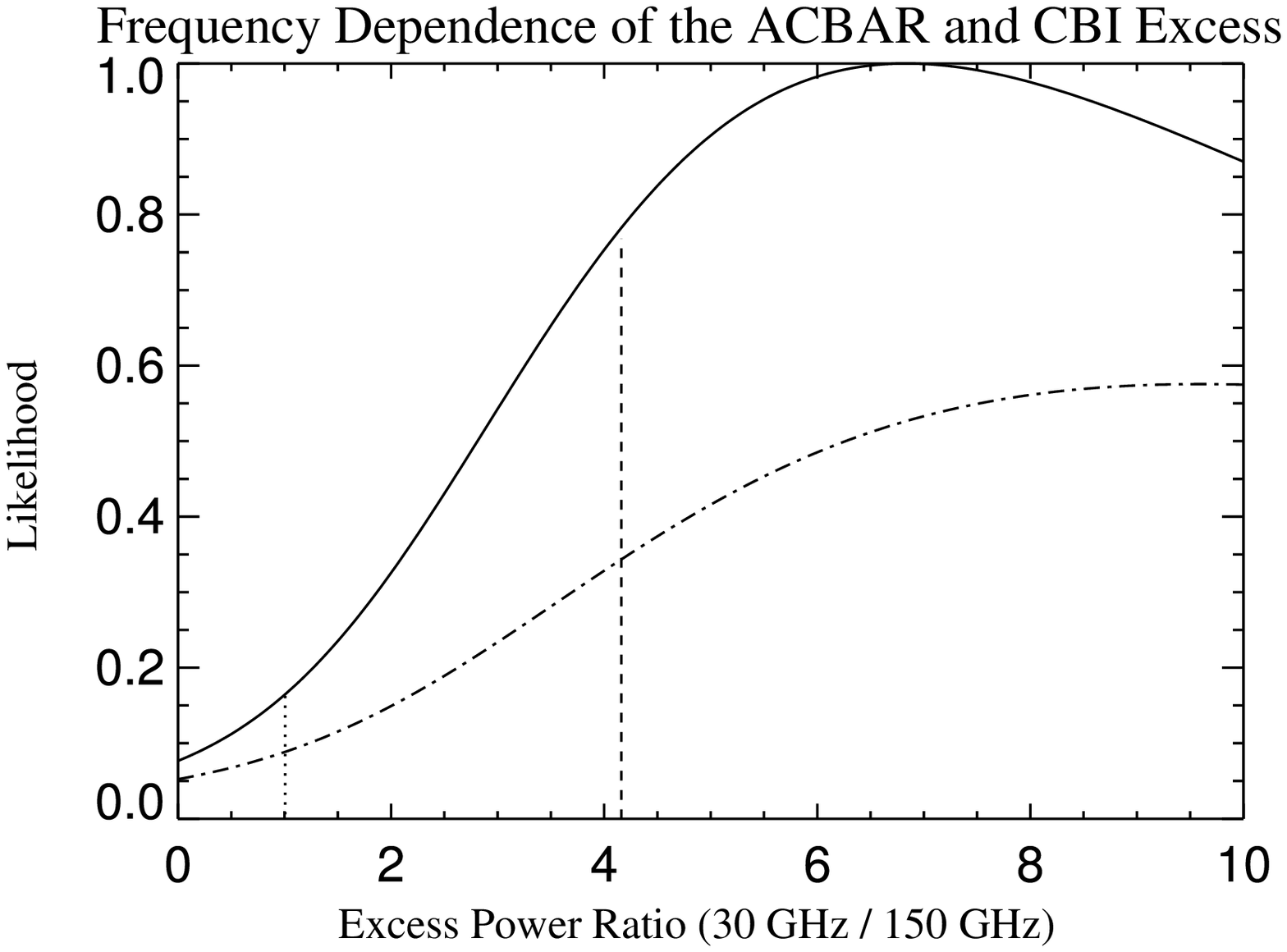}}
\caption{ACBAR results on the high-$\ell$ anisotropies. {\em Left}:
  The ACBAR band-powers above $\ell=1000$ plotted against the best-fit
  ACBAR+WMAP5 model spectrum. The latest CBI results at 30 GHz and the
  previous ACBAR results are also shown.  The ACBAR band-powers for
  $\ell > 1950$ are consistently below the reported high-$\ell$
  CBI band-power.  {\em Right}: The likelihood distribution for
  the ratio of the ``excess" power, observed by CBI at $30\,$GHz and
  ACBAR at $150\,$GHz. The solid line assumes that dusty proto-galaxies do not contribute to the ACBAR band-powers, while the dash-dot line includes the predicted contribution of approximately 22 $\mu {\rm K}^2$ at $\ell = 2600$ from these sources (see \S\ref{subsec:foregrounds}).   With the inclusion of dusty galaxies, the likelihood peaks at a higher power ratio ($\simeq 10$).
The excess power for each experiment is the difference between the 
measured and model band-powers for each experiment in a flat band 
with $\ell>1950$.  
The likelihood for a given power ratio is found from Monte Carlo 
simulations of the band-powers and uncertainties. 
   The vertical dashed line represents the expected ratio
  (4.3) for the excess being due to the SZE, while radio foregrounds 
would correspond to a ratio of $> 10$.  If the excess
  power seen in CBI is caused by non-standard primordial processes,
  the ratio will be unity (blackbody), indicated by the dotted line.
  It is considerably more likely that the excess seen by 
  CBI is caused by either the thermal SZE or radio foreground
  contamination than a primordial source.  }
\label{fig:excess}
\end{figure*}

The ACBAR band-powers reported in this paper are slightly larger at
$\ell > 2000$ than expected for the ``ACBAR+WMAP5" best fit model.  We
subtract the predicted band-powers at $\ell>1950$ from the measured
band-powers in Table~\ref{tab:bands} and find an excess of 
$22 \pm 20 ~\mu K^2$ in a flat band-power from $1950 < \ell < 3000$.  This estimate ignores the band-power contribution from dusty proto-galaxies which is expected to be comparable (see \S\ref{subsec:foregrounds}).
The ACBAR band-powers at $150\,$GHz can be  
used to place constraints on frequency spectrum of the larger CBI 
excess measured at $30\,$GHz. 
We parameterize the excess power for $\ell > 1950$ at the two frequencies as
$P_{30} = \alpha P_{150}$ and sample the likelihood surface for
$\alpha \in [0,10]$ and $P_{150} \in [0 ,300]$ $\mu K^2$.  The beam
uncertainty and the calibration error for both experiments is taken
into account by Monte Carlo techniques.  The likelihood function is
averaged over 1000 realizations under the assumption that each of the
errors has a normal distribution.  The resulting likelihood function for $\alpha$
(after $P_{150}$ is marginalized) is shown in
Figure~\ref{fig:excess}. From the ACBAR and CBI frequency bands, we
expect $\alpha=4.3$ for power originating from the SZE. 
If the excess is due to primary
CMB anisotropies, we expect $\alpha = 1$.  We conclude that it is more than 5
times as likely that the excess seen by CBI and ACBAR is caused by
the thermal SZE than a primordial source.  We expect the
contribution of radio sources to CMB power to be at least a factor of
ten higher at 30 GHz than at 150 GHz ($\alpha \ge 10$).  Because of
the relatively weak detection of excess power by ACBAR, flat spectrum 
radio sources are determined to be $\sim$10\% more likely than the SZE to 
be the source of the excess.
The lower level of excess power seen by ACBAR argues against the 
the CBI excess having a primordial origin, but is consistent 
with either the SZE or radio source foregrounds.  
When we include the expected contribution of dusty protogalaxies to the 
ACBAR excess band-power, the likelihood of the CBI excess being due to radio sources 
increases with respect to thermal SZE.

\section{Cosmological Parameters}\label{sec:parameters}

\subsection{Cosmological Parameters and their ``Prior'' Measures}\label{subsec:basicvars}

In this section, we estimate cosmological parameters for a minimal
inflation-based, spatially-flat, tilted, gravitationally lensed,
$\Lambda$CDM model characterized by six parameters, and then
investigate models with additional parameters to test
extensions of the theory.  For our base model, the six parameters are:
the physical density of baryonic and dark matter, $\Omega_bh^2$ and
$\Omega_ch^2$; a uniform spectral index $n_s$ and amplitude $\ln A_s$
of the primordial power spectrum, the optical depth to last
scattering, $\tau$; and $\theta$, the ratio of the sound horizon at
last scattering to the angular diameter distance.  The primordial
comoving scalar curvature power spectrum is expressed as ${\cal
P}_s(k) = A_s (k/k_\star)^{(n_s-1)}$, where the normalization
(pivot-point) wavenumber is chosen to be $k_\star = 0.05\, {\rm
Mpc}^{-1}$.  The parameter $\theta$ maps angles observed at our
location to comoving spatial scales at recombination; changing
$\theta$ shifts the entire acoustic peak/valley and damping pattern of
the CMB power spectra.  
Additional parameters are derived from this basic
set. These include: the energy density of a cosmological constant in
units of the critical density, $\Omega_\Lambda$; the age of the
universe; the energy density of non-relativistic matter, $\Omega_m$;
the {\it rms} (linear) matter fluctuation level in $8h^{-1}$Mpc
spheres, $\sigma_8$; the redshift to reionization, $z_{re}$; and the
value of the present day Hubble constant, $H_0$, in units of km
s$^{-1}$Mpc$^{-1}$.

Single-field models of inflation predict the existence of a
gravitational wave background characterized by a primordial power law
${\cal P}_t \sim k^{n_t}$. We
characterize the strength by the tensor-to-scalar ratio $r={\cal
P}_t/{\cal P}_s$ evaluated at a pivot point $0.002\, {\rm
Mpc}^{-1}$. We relate the tilt to $r$ using the approximate
consistency relation $n_t \approx -r/8/(1-r/16)$. (We find little
difference in the parameters if we just fix $n_t$ to be zero, as has
often been assumed when $r$ is included, but using this relation is
superior since it is motivated by inflation physics.)

A small running of the spectral index is also expected in slow-roll
inflation and we test for this by extending the basic $\Lambda$CDM
power law model to include a scale dependence of the scalar spectral tilt, 
$d n_s/d\ln (k)$. 

We have also added non-zero curvature $\Omega_k$ to our basic six
parameters. The results are consistent with the flat case, but with the
standard geometrical degeneracy relating $\Omega_k$ and
$\Omega_\Lambda$ expressed through $\theta$ leading to a
near-degenerate tail to $\Omega_k< 0$. 

The ACBAR spectrum includes band-powers at $\ell> 2000$ where the
signal due to secondary CMB anisotropies associated with
post-recombination nonlinear effects should become significant. In
particular, the thermal Sunyaev-Zel'dovich effect and the contribution
of unresolved radio sources and dusty galaxies will ultimately
dominate over the primary anisotropy damping tail; the only question
is at what multipole crossover occurs. Without including such
secondary effects, the parameters we derive from the primary
anisotropy power spectrum could be biased. To account for this, we
have added to the primary anisotropy power spectrum (1) the SZE
template power spectrum ${\hat {\cal D}}_\ell^{\rm SZ}$ used in K07
and \citep{bond05,goldstein03} which was derived from cosmological
hydrodynamics simulations and (2) an unclustered point source
template, as in eq.~\ref{eq:src}.  Each template is scaled by an
overall amplitude parameter, $q_{\rm SZ}$ and $q_{\rm src}$, which we
assume have uniform prior measures with a range much larger than
required by the ACBAR data. The white-noise form for ${\cal D}^{\rm
  src}_\ell$ given by eq.~\ref{eq:src}, is appropriate for the
statistically-averaged power of a distribution of unclustered sources.
The clustering of radio sources is not a large effect, but we do
expect sub-mm sources associated with dusty galaxies at lower flux
levels to be clustered.  As mentioned in \S~\ref{subsec:foregrounds},
in spite of great strides in sub-mm observations in recent years,
significant uncertainties remain in source fluxes and clustering at
$150\,$ GHz.  Theoretical models suggest both will be important for a
complete treatment, but the approximation adopted here should be
sufficient for the ACBAR data set.  In the parameter tables below, we
show results including these secondary templates. We find the basic
parameter central values and uncertainties change little whether we
marginalize over either of the two template amplitudes or set them
both to zero.

The parameter constraints are obtained using a Monte Carlo Markov
Chain (MCMC) sampling of the multi-dimensional likelihood as a
function of model parameters. The pipeline is based on the publicly
available {\textsc CosmoMC}\footnote{http://cosmologist.info/cosmomc}
package \citep{Lewis:2002ah}. CMB angular power spectra and matter
power spectra are computed using the {\textsc CAMB} code
\citep{lewis00}.  As described in Section~\ref{sec:results}, we
approximate the full non-Gaussian band-power likelihoods with an
offset lognormal distribution \citep{bond2000}.  Our standard {\textsc
  CosmoMC} results include the effects of weak gravitational lensing
on the CMB \citep{seljak96,lewis00}. Lensing effects in the
temperature spectrum are expected to become significant at scales
$\ell > 1000$, hence it is important to include this effect when
interpreting the ACBAR results.  The major effect of lensing is a
scale-dependent smoothing of the angular power spectrum which
diminishes the peaks and valleys of the spectrum.  Inclusion of
lensing in the model improves the fit to the data for all experiment
combinations.

The typical computation consists of eight separate chains, each having
different initial, random parameter choices. The chains are run until
the largest eigenvalue of the Gelman-Rubin test is smaller than 0.01
after accounting for burn-in. Uniform priors with very broad
distributions are assumed for the basic parameters. The standard run
also includes a weak prior on the Hubble constant ($45 < H_0 < 90$
km\, s$^{-1}$\, Mpc$^{-1}$) and on the age of the universe ($>10$
Gyrs), but these have negligible effects. We also investigate the
influence of adding Large Scale Structure (LSS) data from the 2 degree
Field Galaxy Redshift Survey (2dFGRS) \citep{cole05} and the Sloan
Digital Sky Survey (SDSS) \citep{tegmark06}. When including the LSS
data, we use only the band-powers for length scales larger than $k
\sim 0.1 ~h~$Mpc$^{-1}$ to avoid non-linear clustering and
scale-dependent galaxy biasing effects. We marginalize over a
parameter $b^2_g$ which describes the (linear) biasing of the
galaxy-galaxy power spectrum for $L_\star$ galaxies relative to the
underlying mass density power spectrum. We adopt a Gaussian prior on
$b^2_g$ centered around $b_g=1.0$ with a very large width. We have
also tried restricting the width to $\delta b_g = 0.3$, but the
cosmic parameters are insensitive to this width.

\subsection {Base Parameter Results}\label{sec:basic}

The results for the basic spatially flat tilted $\Lambda$CDM
parameters are presented in Table~\ref{tab:basic}. The confidence
limits are obtained by marginalizing the multi-dimensional likelihoods
down to one dimension. The median value is obtained by finding the
50\% integral of the resulting likelihood function while the lower and
upper error limits are obtained by finding the 16\% and 84\%
integrals, respectively. The CMBall data combination includes the
ACBAR results presented here and other CMB data sets with published
band-powers and window functions: the WMAP 5 year angular power
spectra \cite{nolta08}, and for comparison the WMAP 3
year spectra \cite{hinshaw06}; the CBI extended mosaic results \citep{readhead04} and
polarization results \citep{Readhead04b,Sievers05}, combined in the
manner described in \citet{Sievers05};\footnote{We exclude the band-powers below $\ell=600$ from the CBI extended mosaic results to
  reduce the correlation with the TT band-powers of the CBI
  polarization dataset which influence the sample-dominated end of the
  spectrum.} the DASI two year results \citep{halverson02}; the DASI
EE and TE band-powers \citep{Leitch04}; the VSA final results
\citep{dickinson04}; the MAXIMA 1998 flight results \citep{hanany00};
and the TT, TE, and EE results from the BOOMERANG 2003 flight
\citep{jones06, piacentini06, montroy06}.  Only $\ell > 350$
band-powers are included for BOOMERANG because of overlap with WMAP
(although inclusion of the lower $\ell$ results leaves the parameter
results essentially unchanged).  While ACBAR and BOOMERANG are both
calibrated through WMAP, this is a small contribution to the total
uncertainty in the ACBAR calibration and we treat the calibration
uncertainties as independent in our parameter analysis. Although the
DASI, CBI, and BOOMERANG 2003 EE and TE results for high-$\ell$
polarization are included, they have little impact on the values of
the parameters we obtain. 

\begin{figure*}[th!]
\resizebox{\hsize}{!}{
\plotone{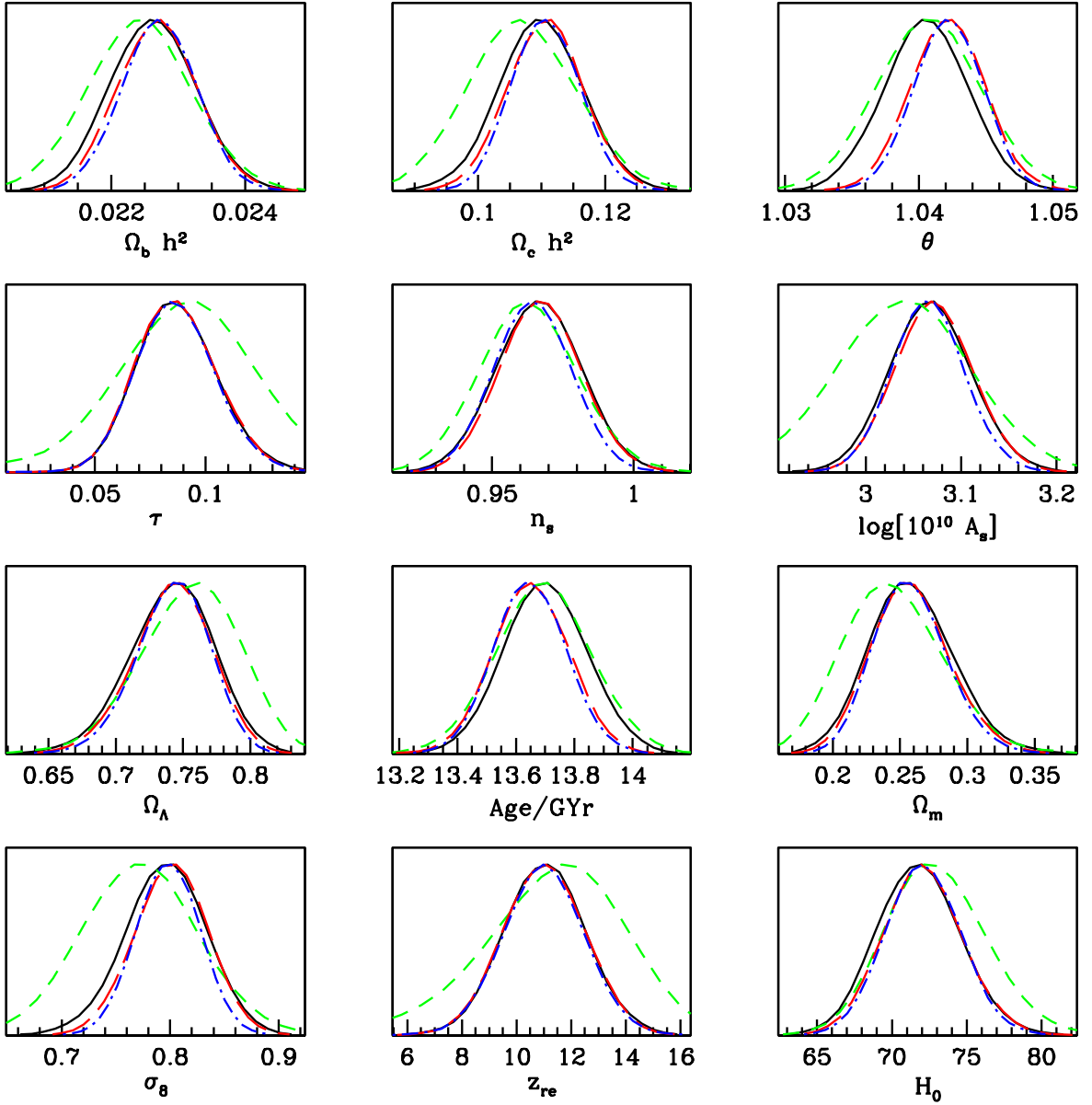}}
\caption{Basic parameter marginalized 1-dimensional likelihood
  distributions for the following data combinations; WMAP3-only
  (green, dashed), WMAP5-only (black, solid), ACBAR + WMAP5 (red,
  long-dashed), CMBall (blue, dot-dashed). All runs include
  lensing.}
\label{fig:basic}
\end{figure*}

The latest WMAP likelihood code found at http://lambda.gsfc.nasa.gov/
has been used in our analyses. When this ACBAR paper was submitted,
all parameter analyses were done using WMAP3 \citep{hinshaw06}; 
the new WMAP5 results came out a few months later. 
The WMAP5 data allowed for an improved cross calibration of the ACBAR
and WMAP band-powers and resolved the ($<$ 1-$\sigma$) parameter tensions 
that existed between the WMAP3 and ACBAR results.
We note that the WMAP5 team's parameter analysis \citep{dunkley08} 
made use of the ACBAR band-powers with the $1.4\%$ higher temperature calibration 
from the WMAP3 and ACBAR cross-calibration. 
The marginalized one-dimensional likelihood
distributions for the basic parameter set we obtain are shown in
Figure~\ref{fig:basic}. Note the contrast between the parameter
determinations for WMAP3 and WMAP5. The primary improvement of WMAP5
over WMAP3 was a better understanding of the beam, which resulted in an 
improved measurement of the third acoustic peak. 
Other improvements in WMAP5 were updated point-source
correction, stimulated by \citep{huffenberger06}, and foreground
marginalization on large angular scales.

\begin{figure*}[th!]
\resizebox{\hsize}{!}{
\plotone{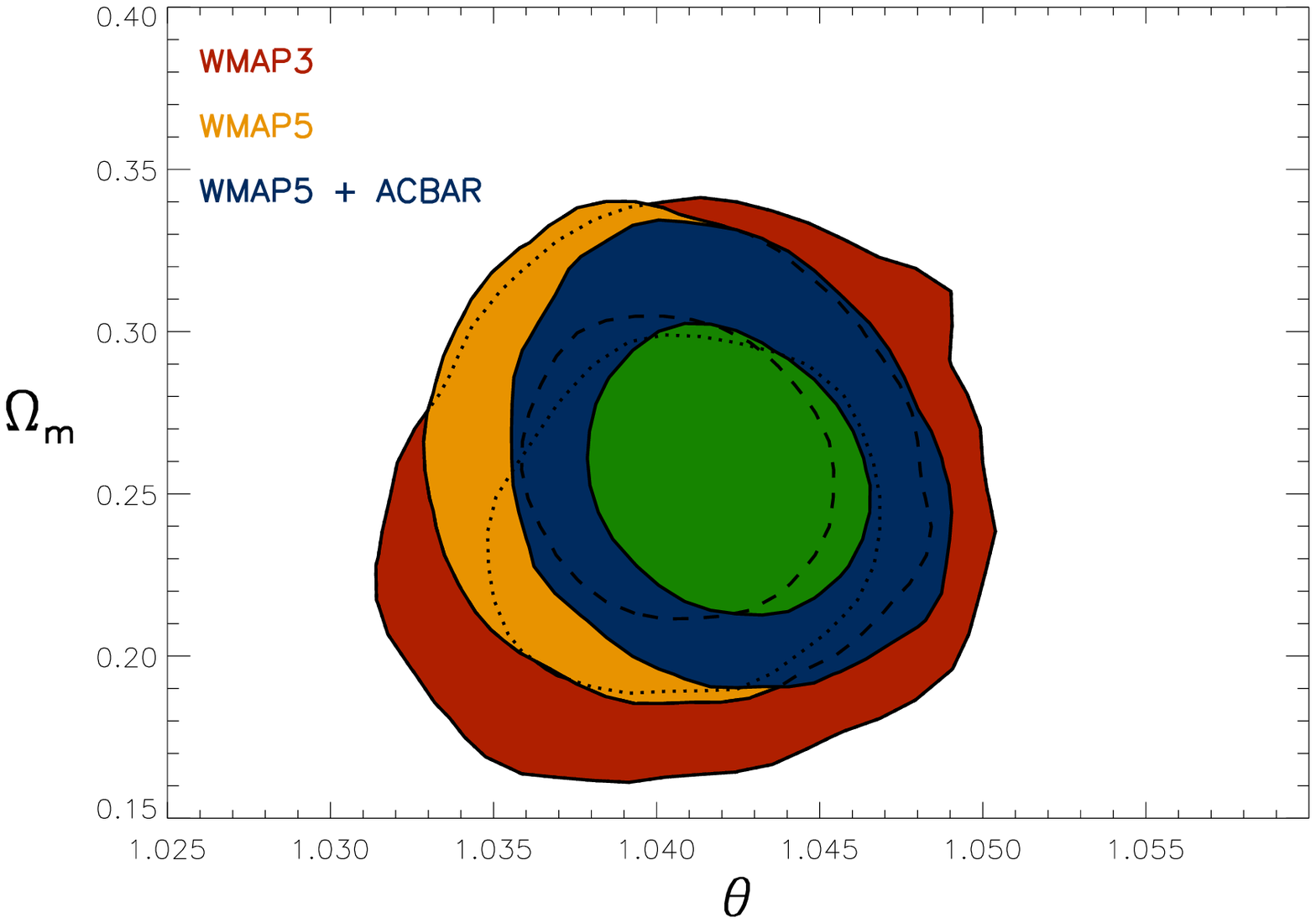}}
\caption{68\% and 95\% 2D marginalized contours in $\theta$ and
  $\Omega_m$ for a number of data combinations. The results for $\theta$
from WMAP are largely driven by the determination of the third peak. 
The value preferred by the WMAP5 data is slightly higher than that from WMAP3,
and more consistent with that found when ACBAR is added to either WMAP data
set.}
\label{fig:theta}
\end{figure*}
The addition of LSS data has little impact on the mean values and
errors in the cosmic parameters. The largest shift is $<$ 1-$\sigma$ in
$\Omega_ch^2$, from $0.111^{+0.005}_{-0.005}$ to
$0.107^{+0.004}_{-0.004}$. Our LSS results only include information on
the shape of the density power spectrum, not its overall amplitude
since we marginalize over the galaxy bias factor. If results from weak
lensing are included in the LSS data, then there is a slight increase
in $\sigma_8$, but it depends somewhat on which
lensing results are included. The effect is to slightly increase the
CMB+LSS result and improve the consistency with the CMBall results.
The third peak is well determined by both ACBAR and WMAP5, and this defines 
the dark matter density;
the inclusion of the LSS data does not significantly improve the constraints.  
We have also found that the results do not change significantly from
CMBall+LSS when SN1a data are included (in this case from the
\citet{riess04} gold set), so we have not included a separate
column. Including SN1a data would be crucial if we were attempting to constrain 
the equation of state of dark energy.

\begin{figure*}[th!]
\resizebox{\hsize}{!}{
\plotone{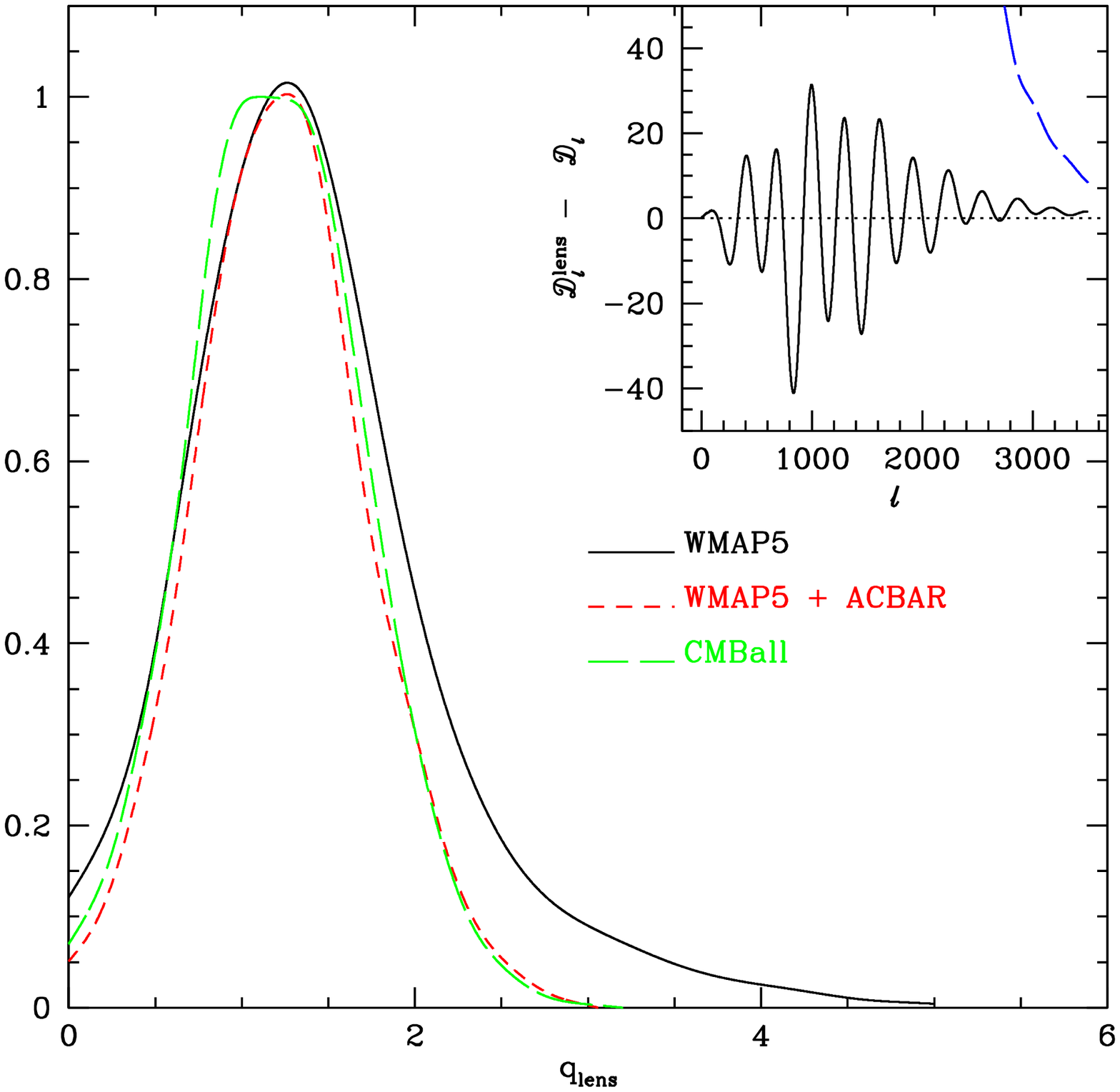}}
\caption{The 1-dimensional marginalized likelihood distribution for the lensing
  amplitude $q_{\rm lens}$ for the combinations of CMB data
  shown. Although only $q_{\rm lens}=1$ is physically meaningful,
  these distributions indicate that the data prefer lensing of about 
the right magnitude. The ratio of the value 
  at $q_{\rm lens}=1$ to that at $q_{\rm lens}=0$ is the
  Bayesian evidence that lensing at the right amplitude is preferred
  over no-lensing. The inset figure shows $\Delta {\cal D}_\ell^{\rm
    lens} \equiv {\cal D}_\ell^{\rm lens} - {\cal D}_\ell^{\rm
    no-lens}$, with both spectra computed using the best-fit parameters 
  form the lensed analysis of the 
  WMAP5+ACBAR data. Notice how small this correction is relative to
  ${\cal D}_\ell^{\rm lens}$, shown as the dashed blue curve. 
The cosmological parameters used to construct $\Delta {\cal D}_\ell^{\rm
    lens}$ have been allowed to vary as the data requires in
  constructing the marginalized likelihoods for $q_{\rm lens}$. Each
  parameter variation leads to a slightly different difference
  template than that shown in the inset. }
\label{fig:lensfit}
\end{figure*}

All parameter results listed in Tables~\ref{tab:basic} and
\ref{tab:nrun} include the effects of weak lensing of the CMB on the
resulting power spectrum.  For every case including CMB lensing, we
have performed an identical calculation neglecting the effects of
lensing.  Including lensing improves the fit of the model to the
observed band-powers compared for all data combinations. 
This can be quantified by the log-ratio of the lensed to
no-lensed Bayesian evidence, $\Delta \ln {\cal E}=\ln[P({\rm
  lens}|data,theory)/P({\rm no-lens}|data,theory)]$. The evidence
$P({\rm lens}|data, theory)$ is an integral of the product of the {\it
  a priori} probability (the parameters' measure) and the likelihood
of data given those parameters; it appears in the denominator in the
Bayesian chain to ensure the {\it a posteriori} probability has unity
normalization. The resulting number is a conditional probability given
the data and the assumptions about the parameters.
The parameters and their measures are
exactly the same, so the ratio is a robust indicator of preference. For
WMAP5 alone it is $\Delta \ln {\cal E}=2.04$; it increases to 2.89
with ACBAR included; and is 2.63 for CMBall. Naively relating this to
a Gaussian translates to a significance of $\sim 2.3 \sigma$ for CMBall.  From
Fig.~\ref{fig:acbar}, it is clear that the difference in the
power spectra of lensing and no-lensing is small; for this plot, the 
best-fit parameters from the lensed analysis were fixed and used to 
compute a lensed and unlensed spectrum.  
The difference, $\Delta
C_\ell^{\rm lens} \equiv C_\ell^{\rm lens} - C_\ell^{\rm no-lens}$, is
explicitly shown in the inset of Figure~\ref{fig:lensfit}. We find
only small shifts in the median value of the cosmic parameters when
lensing is included; e.g., for the ACBAR+WMAP5 data combination, we
find $\sigma_8=0.79^{+0.03}_{-0.03} \rightarrow 0.80^{+0.03}_{-0.03}$
and $\Omega_ch^2 = 0.109^{+0.006}_{-0.006}\rightarrow
0.111^{+0.006}_{-0.006}$ when going from non-lensed to lensed models
respectively.

\begin{deluxetable*}{c||cccccc}
\tabletypesize{\scriptsize}
\tablecaption{ Basic 6 Parameter Constraints } 
\setlength{\tabcolsep}{0.04in} 
\tablehead{ \colhead{} & \colhead{WMAP5} & \colhead{WMAP5+ACBAR}  & \colhead{CMBall} & \colhead{CMBall+LSS} & \colhead{CMBall+$q_{\rm SZ}$} &\colhead{CMBall+SZ+$q_{\rm src}$}}

\startdata
$        \Omega_bh^2$  & $0.0226^{+0.0006}_{-0.0006}  $&$0.0227^{+0.0006}_{-0.0006}  $&$0.0227^{+0.0005}_{-0.0005} $&$ 0.0228^{+0.0005}_{-0.0005} $&$0.0227^{+0.0006}_{-0.0005} $&$ 0.0227^{+0.0005}_{-0.0005}  $     \\
$        \Omega_ch^2$  & $0.110^{+0.006}_{-0.006}     $&$0.111^{+0.006}_{-0.006}     $&$0.111^{+0.005}_{-0.005}    $&$ 0.107^{+0.004}_{-0.004}    $&$0.109^{+0.005}_{-0.005}    $&$ 0.111^{+0.005}_{-0.005}     $          \\
$          \theta$     & $1.041^{+0.003}_{-0.003}     $&$1.042^{+0.003}_{-0.003}     $&$1.042^{+0.002}_{-0.002}    $&$ 1.042^{+0.002}_{-0.002}    $&$1.042^{+0.002}_{-0.002}    $&$ 1.042^{+0.002}_{-0.002}     $          \\
$            \tau$     & $0.086^{+0.008}_{-0.008}     $&$0.086^{+0.008}_{-0.008}     $&$0.086^{+0.008}_{-0.008}    $&$ 0.087^{+0.008}_{-0.008}    $&$0.086^{+0.008}_{-0.008}    $&$ 0.085^{+0.008}_{-0.008}     $          \\
$             n_s$     & $0.967^{+0.015}_{-0.015}     $&$0.967^{+0.014}_{-0.014}     $&$0.964^{+0.013}_{-0.013}    $&$ 0.968^{+0.012}_{-0.012}    $&$0.962^{+0.013}_{-0.013}    $&$ 0.962^{+0.013}_{-0.013}     $  \\
$     {\cal A}_s$      & $3.07^{+0.04}_{-0.04}        $&$3.07^{+0.04}_{-0.04}        $&$3.07^{+0.04}_{-0.04}       $&$ 3.05^{+0.04}_{-0.04}       $&$3.05^{+0.04}_{-0.04}       $&$ 3.06^{+0.04}_{-0.03}        $     \\      
$  \Omega_\Lambda$     & $0.74^{+0.03}_{-0.03}        $&$0.74^{+0.03}_{-0.03}        $&$0.74^{+0.02}_{-0.03}       $&$ 0.76^{+0.02}_{-0.02}       $&$0.75^{+0.02}_{-0.03}       $&$ 0.74^{+0.02}_{-0.03}        $     \\ 
$             Age$     & $13.7^{+ 0.1}_{- 0.1}        $&$13.7^{+ 0.1}_{- 0.1}        $&$13.6^{+ 0.1}_{- 0.1}       $&$ 13.6^{+ 0.1}_{- 0.1}       $&$13.7^{+ 0.1}_{- 0.1}       $&$ 13.7^{+ 0.1}_{- 0.1}        $     \\ 
$        \Omega_m$     & $0.26^{+0.03}_{-0.03}        $&$0.26^{+0.03}_{-0.03}        $&$0.26^{+0.03}_{-0.02}       $&$ 0.24^{+0.02}_{-0.02}       $&$0.25^{+0.03}_{-0.02}       $&$ 0.26^{+0.03}_{-0.02}        $     \\ 
$        \sigma_8$     & $0.80^{+0.04}_{-0.04}        $&$0.80^{+0.03}_{-0.03}        $&$0.80^{+0.03}_{-0.03}       $&$ 0.78^{+0.03}_{-0.02}       $&$0.79^{+0.03}_{-0.03}       $&$ 0.80^{+0.03}_{-0.03}        $     \\ 
$          z_{re}$     & $11.0^{+ 1.4}_{- 1.4}        $&$11.0^{+ 1.5}_{- 1.4}        $&$11.0^{+ 1.4}_{- 1.4}       $&$ 11.0^{+ 1.4}_{- 1.4}       $&$10.9^{+ 1.4}_{- 1.4}       $&$ 10.9^{+ 1.4}_{- 1.4}        $     \\ 
$             H_0$     & $71.8^{+ 2.7}_{- 2.7}        $&$72.0^{+ 2.6}_{- 2.5}        $&$72.1^{+ 2.4}_{- 2.4}       $&$ 73.6^{+ 1.9}_{- 1.9}       $&$72.5^{+ 2.4}_{- 2.4}       $&$ 72.0^{+ 2.4}_{- 2.3}        $      \\   
$     q_{\rm SZ}$ &\nodata&\nodata&\nodata&\nodata&                                                                                                 $0.69^{+0.11}_{-0.11}       $   &\nodata                             \\
$    q_{\rm src}$ &\nodata&\nodata&\nodata&\nodata&\nodata&                                                                                                                     $   29^{+12}_{-28}               $      \\
$ \sigma_8^{{\rm SZ}}$ &\nodata&\nodata&\nodata&\nodata&                                                                                            $ 0.93^{+0.04}_{-0.05}      $   &\nodata                        \\
\enddata
\tablecomments{\small
Results for the basic parameter set. The runs all
  assume flat cosmologies, uniform and broad priors on each of the
  basic six parameters, and a weak prior on the Hubble constant ($45 <
  H_0 < 90$ km\, s$^{-1}$ Mpc$^{-1}$) and the age of the universe ($> 10$ Gyr). Here ${\cal
    A}_s\equiv \log[10^{10}A_s]$. All runs include the effect of weak
  gravitational lensing on the CMB. Column 5 presents the results when
  a SZ template characterized by the overall SZ-template-power
  $q_{\rm SZ}$ is included. 
  The values of $\sigma_8^{{\rm SZ}}= q_{\rm
  SZ}^{1/7} (\Omega_bh)^{2/7}$ are higher than the $\sigma_8$
  derived from the primary anisotropies. Column 6 shows the results
  when the point source contribution scaled by $q_{\rm src}$ is
  included in combination with the SZ-template with $q_{\rm SZ}$ set to
  $\sigma_8^7(\Omega_bh)^2$. 
  The marginalization over either extra high-$\ell$ contribution does not 
  significantly shift the results for the basic parameters. }
\label{tab:basic}
\end{deluxetable*}


We now test whether the strength of
the lensing modification is consistent with expectations for lensing,
by multiplying the lensing template $\Delta
C_\ell^{\rm lens}$, which varies with cosmic parameters, by a
variable strength $q_{\rm lens}$:
\begin{equation} C_\ell^{\rm lens} = C_\ell^{\rm no-lens} + q_{\rm
  lens} \Delta C_\ell^{\rm lens} \, .
\end{equation} 
The normalization is such that $q_{\rm lens}=1$ gives the normal
lensed CMB spectrum, while $q_{\rm lens}=0$ gives the no-lensing case.
An accurate determination of this subtle effect requires
highly accurate window functions for the bands. We use a flat prior
probability for $q_{\rm lens}$, allowing it to vary from 0 to 10.
With WMAP5 alone, we obtain $q_{\rm lens}=1.34^{+0.27 (+1.51)}_{-0.26
  (-0.85)}$; WMAP5+ACBAR gives $q_{\rm lens}=1.23^{+0.21
  (+0.83)}_{-0.23 (-0.76)}$; CMBall gives $q_{\rm lens}=1.21^{+0.24
  (+0.82)}_{-0.24 (-0.76)}$. We have also listed the 2-$\sigma$ errors
(in brackets), which are far from twice the 1-sigma values, reflecting
the highly non-Gaussian nature of the marginalized likelihoods evident
in the figure. Although we emphasize that $q_{\rm lens}$ is not in any
sense an independent parameter, it does illustrate that lensing 
of the expected strength is preferred.\footnote{\cite{slosar08} also undertook a lensing analysis of
  ACBAR temperature power spectrum, but, instead of $q_{\rm lens}$ as
  defined here, they used a multiplier $A_{\rm L}$ of the lensing
  potential power spectrum, defined to be unity for normal lensing;
  they found $A_{\rm L}=3.0^{+0.9}_{-0.9}$ for WMAP5+ACBAR. Repeating our 
analysis with this parameterization, we find lower values, $A_{\rm
    L}=1.60^{+0.55 (+1.79) }_{-0.26 (-0.99)}$. With the highest 3 ACBAR bins cut out,
where secondary effects might have an impact on the determination of
the lensing strength parameter, the results are essentially unchanged. }
The significance of the detection is somewhat less than the WMAP3/NVSS/SDSS 
cross-correlation results of \citet{smith07} and \citet{hirata08}. In
Figure~\ref{fig:lensfit}, we show the marginalized distribution for
the $q_{\rm lens}$ parameter using various combinations of CMB data.

We have also run a limited set of non-flat model chains. The models in
this case do not include the effect of weak lensing and we keep the
same weak prior on $H_0$. When $\Omega_k=0$ is not enforced, the weak
prior on $H_0$ has a significant effect on the result as it restricts
the extent of the geometrical degeneracy which is present in this
case. For WMAP5 only we obtain $\Omega_k=-0.018^{+0.027}_{-0.026}$ which
becomes $\Omega_k=-0.013^{+0.019}_{-0.029}$ when ACBAR is included.
\subsection{Residual source marginalization}\label{subsec:ressourcemarg}

As discussed in \S~\ref{subsec:foregrounds}, the ACBAR band-powers at
$\ell>1950$ marginally exceed the predictions of the best-fit models
for the primary CMB.  
We have repeated the basic parameter runs including an unclustered source 
contribution described by equation~\ref{eq:src} and marginalize over a wide uniform
prior in $q_{\rm src}$ from 0 to $4600~\mu K^2$, i.e. more than 100
times the power required to fit the high-$\ell$ points.  
The results for runs including this marginalization are shown in
Tables~\ref{tab:basic} and~\ref{tab:nrun}.  
The template amplitude $q_{\rm src}$ is not well constrained by the fit,
and the effect of the marginalization on the basic parameters is
negligible.  The addition of a source template improves the fit to the
high-$\ell$ points and therefore increases, albeit modestly, the
best-fit likelihood: for the ACBAR+WMAP5 combination the change in
likelihood is $\Delta \ln L=0.24$.
The value and uncertainty for $q_{src}$ given in Table~\ref{tab:basic} spans the range of 
predictions for the contribution from dusty protogalaxies. 

Contributions from point sources or the SZE are nearly
degenerate in the high-$\ell$ ACBAR band-powers. However, the CMBall
combination is potentially sensitive to an SZE contribution because of
the SZE frequency dependence. The CBI data has a more significant excess
at high-$\ell$, however, through substantial modeling and vetoing, the CBI band-powers
are expected to be free of radio source contamination. 
We assume that the residual contribution of the unresolved source background to the CBI
band-powers is sufficiently low that we do not require a source
template for that data, only the SZE template.

\subsection{Sunyaev-Zel'dovich template extension}\label{sec:sz}

The amplitude of the SZE signal depends strongly on the overall matter
fluctuation amplitude, $\sigma_8$.  We have modified our parameter
fitting pipeline to allow for extra frequency dependent contributions
to the CMB power spectrum and have implemented it in a simple analysis
using a fixed template ${\hat{\cal D}}_\ell^{\rm SZ}$ for the shape of
the thermal SZE power spectrum. The template was obtained from two
large, hydrodynamical simulations of a scale-invariant ($n_s=1$)
$\Lambda$CDM model with $\Omega_bh = 0.029$ and with $\sigma_8=0.9$
and $\sigma_8=1.0$ . (See \cite{bond05} for a detailed description of
the simulations.)  Recently the WMAP team have used a different SZE
template based on analytic estimations of the power spectrum
\citep{spergel06}. It is characterized by a slower rise in $\ell$ than
the simulation-based template, which cut nearby clusters out of the
power spectrum. There has been no fine-tuning of either spectra to
agree with all of the X-ray and other cluster data. This may have an
effect on shape, especially at high-$\ell$.

The SZE contribution, ${\cal D}_\ell^{\rm SZ} = (q_{\rm SZ})f_\nu {
  \hat{\cal D}}_\ell^{\rm SZ}$, added to the base six-parameter model
spectrum has a frequency-dependent SZE pre-factor $f_\nu$. Including
this SZE template with all model parameters free to vary is
complementary to the analysis of \S~\ref{sec:excess} which directly
compared the residual CBI and ACBAR band-powers at $\ell > 1950$ for
the best fit WMAP5+ACBAR model power spectrum. In that more
restrictive analysis, the primary power spectrum is fixed and $f_\nu$
is allowed to vary as well as a broad-band excess power.  We found the
ratio of excess power seen by CBI and ACBAR to be compatible with 
the ratio of the frequency prefactors $f_\nu$ at 30 GHz and 150 GHz
for the Sunyaev-Zel'dovich effect, although foreground contamination of 
both experiments could also contribute to the observed excess. 
In this section, we assume that all the excess power seen by the CBI experiment 
is due to the SZE.
In the hydrodynamical simulations used to derive the SZE template, the
amplitude was shown to scale as $q_{\rm SZ} = (\sigma_8/0.9)^{7}
(\Omega_bh/0.029)^2$.  We consider two cases: (1) the scaling
parameter $q_{\rm SZ}$ is slaved to the results from the hydrodynamical 
simulations, which means that it is primarily determined by
the primary CMB data; (2) $q_{\rm SZ}$ is allowed to float freely and
an independent $\sigma_8^{\rm SZ}$ is derived, to be compared with the
$\sigma_8$ that is derived from the basic six parameters. We use a
uniform prior in $q_{\rm SZ}$ in this case with limits $0\le q_{\rm
  SZ}\le 4.0$.

Regardless of the data combination, we find that including an SZE
component in the model has little effect on the values of most basic
cosmological parameters (see Table~\ref{tab:basic}), whether $q_{\rm
  SZ}$ is related to cosmic parameters through $q_{\rm SZ} =
(\sigma_8/0.9)^{7} (\Omega_bh/0.029)^2$ or is allowed to float freely.
Note that the SZE results break the $A_s e^{-2\tau}$ near-degeneracy
(as does weak lensing, though not as strongly).

\begin{figure*}[ht!]
\centering
\includegraphics[width=5.5in]{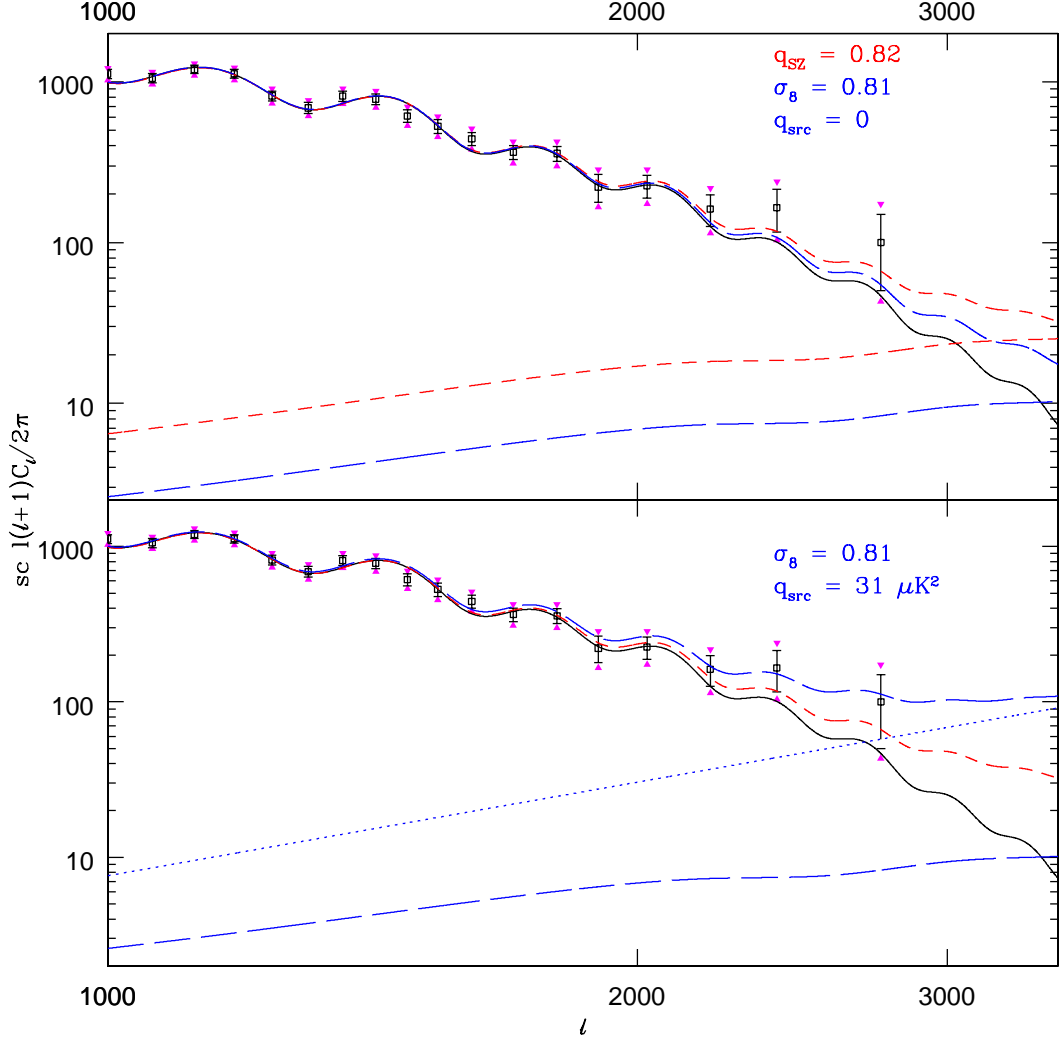}
\caption{ Best-fit models for the ACBAR + WMAP5 combination. Only the
  $\ell>1000$ ACBAR band-powers are shown. The arrows are indicative of
  the possible (coherent) shift in the 1-$\sigma$ confidence limits
  due to beam error. The top panel shows fits with just primary CMB
  (black, solid) and with an extra SZE contribution. The case where the
  SZE template is scaled independently as $q_{\rm SZ}$ gives the best
  fit to the high-$\ell$ excess (red, short-dashed) with $q_{\rm
    SZ}=0.82$. The case where the SZE amplitude is slaved to the
  cosmological parameters $\sigma_8^7(\Omega_bh)^2$ (blue,
  long-dashed) does not yield enough power to fit the excess given the
  best-fit value of $\sigma_8=0.81$. The bottom panel shows the case
  when a point source contribution scaled by $q_{\rm src}$ is included
  for ACBAR together with the slaved SZE contribution (blue,
  dotted). In this case the best-fit model has $\sigma_8=0.81$
  determining the sub-dominant SZE contribution and $q_{\rm src}=31\mu
  K^2$ determining the point source contribution.}
\label{fig:best_fit}
\end{figure*}

With the combination of the ACBAR and WMAP5 data, for which there is only a weak
indication of excess power, we
find a freely-floating SZE amplitude results in $q_{\rm
  SZ}=0.94^{+0.35}_{-0.93}$, with no effective lower bound. We can use
the above relation of $q_{\rm SZ}(\sigma_{8},\Omega_b h)$ to estimate
a corresponding $\sigma_{8}^{({\rm SZ})}= 0.97^{+0.09 (+0.13)}_{-0.13
  (-0.31)}$, where in brackets we have indicated the $2\sigma$ errors.
This result is higher than, but compatible within the
uncertainties, to values obtained from the primary CMB fits alone: the
ACBAR + WMAP5 fits in Table~\ref{tab:basic} give $\sigma_8 =
0.80^{+0.03}_{-0.03}$. The confidence limits of the derived
$\sigma_{8}^{({\rm SZ})}$ depend strongly on the choice of measure,
which is here taken to be uniform in the amplitude $q_{\rm SZ}$ (This
is evident in the translation of the relative flatness of the
likelihood at low $q_{\rm SZ}$ to a ``1-sigma detection'' in
$\sigma_8^{({\rm SZ})}$). When the SZE contribution is slaved to the
$\sigma_8$ and $\Omega_bh$ values from the primary spectrum, there is no effect
on $\sigma_8$. 
The high-$\ell$ excess power is not significant
enough to change the well constrained value of $\sigma_8$ to the weakly preferred
higher value.
The effect of slaved and unslaved fits can be seen in
Fig.~\ref{fig:best_fit}.

When the high-$\ell$ band-powers of CBI and BIMA are included in the
analysis, there is a significant detection of excess power. Both the
CBI and BIMA band-powers are from $30\,$GHz interferometric
observations and have higher $f_\nu$ values than ACBAR. For the slaved
case the $\sigma_{8}$ value and errors are unchanged. For the floating
case, we find $q_{\rm SZ}= 0.69^{+0.11}_{-0.11}$ which maps to
$\sigma_{8}^{({\rm SZ})}=0.93^{+0.04}_{-0.05}$.

These central values and uncertainties are computed by transforming
integrals of the likelihood $L(q_{\rm SZ}, \Omega_bh , ... )$ over the
prior measures to confidence limits for $\sigma_{8}^{(SZ)}$. Exactly
what measure to place on $q_{\rm SZ}$ and therefore on
$\sigma_{8}^{(SZ)}$ is debatable. A measure uniform in
$\alpha_{SZ}=q_{\rm SZ}^{1/2}$, as we used in K07 and
\citep{goldstein03} translates into a measure $\propto q_{\rm
  SZ}^{-1/2}dq_{\rm SZ}$ which favors lower values of
$\sigma_{8}^{(SZ)}$ than the measure uniform in $q_{SZ}$ that we have
adopted.

As the cosmological parameters vary, the SZE template may depend on
$\sigma_{8}$ and $\Omega_b h$, and certainly depends on the spectral
index $n_s$ and astrophysical issues such as the history of energy
injection into the cluster system. In a more complete treatment than
that presented here, the shape should be modified along with the base
cosmological parameters in the MCMC runs.

We caution that the derived $\sigma_{8}^{({\rm SZ})}$ depends on the
SZE template shape, its extension into the higher $\ell$ regime probed
by BIMA, and the prior measure placed upon $q_{\rm SZ}$.\footnote{We
  also note that the non-Gaussian nature of the SZE signal was included
  in the BIMA results, but not in the CBI results.  The non-Gaussian
  effect increases the sample variance and tends to open up the
  allowed range towards lower $\sigma_8$ values
  \citep{goldstein03,readhead04}.} The modest decrease in log likelihood for
the fit to the model when the SZE is taken into account, is $\Delta \ln
L = 0.26$ for the ACBAR+WMAP5 combination and increases to $\Delta \ln
L = 1.91$ for CMBall+BIMA.

\begin{figure*}[ht!]
\resizebox{\hsize}{!}{
\plotone{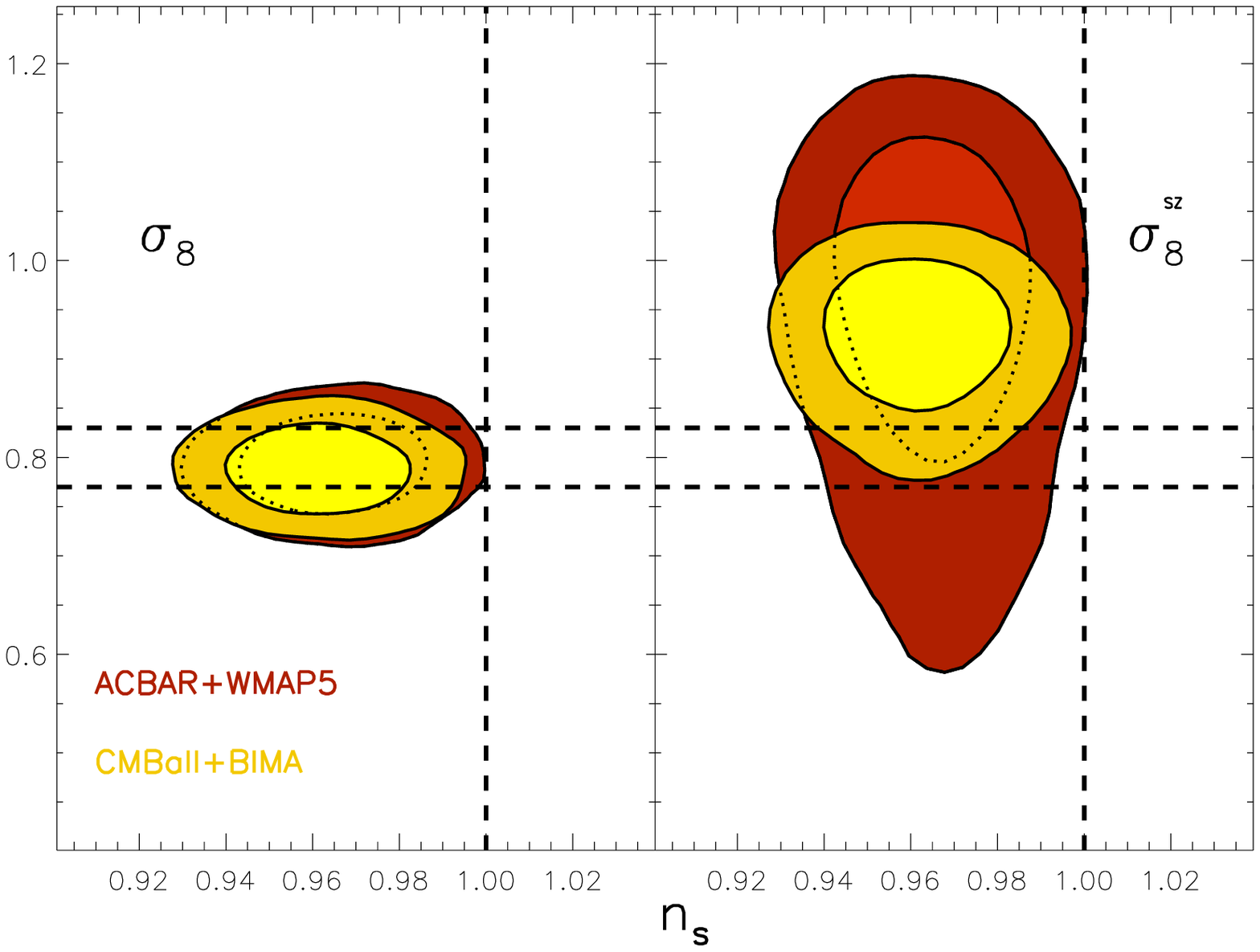}}
\caption{The figure contrasts the one and two sigma contour intervals
  for $\sigma_8$ determined from the primary anisotropy component of
  the CMB (left) with the value inferred from the SZE template
  transformation of $q_{\rm SZ}$ into $\sigma_{8}^{({\rm SZ})}$
  (right), assuming a uniform prior measure in $q_{\rm SZ}$. Allowing
for a point source contribution would decrease the tension between $\sigma_8$ 
and $\sigma_8^{SZ}$ for the ACBAR+WMAP5 case. These 
  panels also demonstrate the strength of the deviation of $n_s$
  from unity for the flat $\Lambda$CDM model.  }
\label{fig:SZ}
\end{figure*}

Regardless of the assumed prior, the interpretation of excess power
as being due to the SZE results in a non-zero $\sigma_{8}^{(SZ)}$ 
for the CMBall+BIMA combination.
Uncertainties for $\sigma_{8}^{({\rm SZ})}$ are about a factor of two
larger than for the $\sigma_8$ determined from the primary CMB data
and there is a tension at about the 2-sigma level between the two
median values.  A visual summary of the results is shown in
Fig.~\ref{fig:SZ} where we plot both $\sigma_{8}$ and
$\sigma_{8}^{({\rm SZ})}$ against the spectral index for a number of
data combinations.  The addition of LSS data does not significantly
change these results.  When the ACBAR dusty point source or SZE template
marginalization is included, $\sigma_8$ decreases slightly for the
CMBall dataset.

Recent weak lensing results are in basic agreement with the primary
$\sigma_8$ values when $\Omega_m = 0.26 \pm 0.03$ from
Table~\ref{tab:basic} is used.  With 100 deg$^2$ of lensing data from
the combination of the CFHT weak lensing legacy, RCS, Virmos-Descart
and GaBaDos surveys, \cite{benjamin07} get $\sigma_8
(\Omega_m/0.26)^{0.59} = 0.80 \pm 0.05$.  From the CFHT weak lensing
legacy survey alone, \cite{fu07} get $\sigma_8 (\Omega_m/0.26)^{0.64}
= 0.753 \pm 0.043$, and get $\sigma_8 (\Omega_m/0.26)^{0.53} = 0.82
\pm 0.084$ if only large-scale linear-regime results are used. These
weak lensing numbers are lower than past published results because of
improved treatments of the redshift distribution of the lensed
sources. 

As shown in Fig.~\ref{fig:best_fit}, the marginal excess power in ACBAR
is consistent with the combination of a point source contribution at the upper limit 
of the $150\,$GHz extrapolations and an SZE template at the level predicted from the 
primary anisotropy $\sigma_8$ value. In this scenario, the cosmological
parameters are virtually unchanged, but no explanation is provided for the
CBI excess. The SZE
analyses with a free-floating amplitude have not included additional
foreground sources for CBI, BIMA, or ACBAR. The effect of radio
sources extrapolated to 30~GHz was included in the original CBI and BIMA
results, and is unlikely to be an important contaminant for
ACBAR. Dusty proto-galaxies should not effect CBI and BIMA, but will
contribute to the high-$\ell$ ACBAR band-powers. 
Given the uncertainty in the source contamination
for ACBAR, the weak detection of excess power 
does not significantly support the SZE interpretation of the CBI+BIMA excess.

\subsection{Running Spectral Index}\label{sec:nrun}

\begin{figure*}[th!]
\resizebox{\hsize}{!}{
\plotone{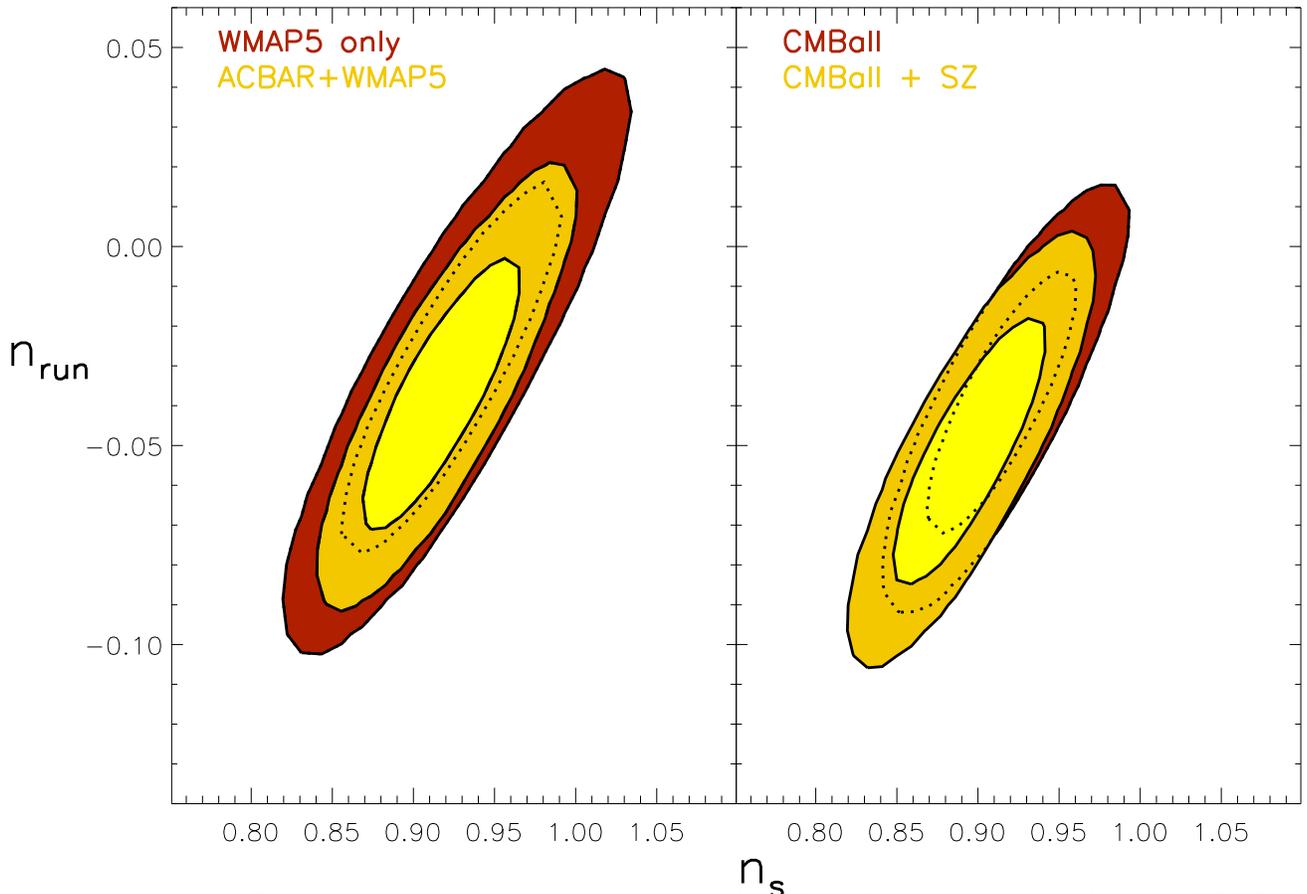}}
\caption{68\% and 95\% 2D marginalized contours in $n_s$ and $n_{run}
  =dn_s/d\ln k$. The left panel shows the results for the WMAP5 and
  ACBAR+WMAP5 combinations.  The right panel shows the effect of
  adding the rest of the CMB data and marginalizing over the SZE
  template.  The basic parameters are essentially unchanged 
  if we marginalize over one or the other of the two template
  amplitudes.}
\label{fig:nrun}
\end{figure*}

There has been interest in the running of the spectral index
$dn_s/d\ln k$ since the first release of WMAP data, which showed
evidence for a significant negative running when combined with LSS and
Lyman alpha forest observations \citep{spergel03}. With the precision
of the new ACBAR data, we might expect improved constraints on running
of the spectral index. To the basic six parameters in the minimal
model, we add running of the spectral index $dn_s/d\ln k (k_\star)$
around the pivot point $k_\star=0.05$ Mpc$^{-1}$.  We adopt the
conventionally-used uniform prior in $dn_s/d\ln k (k_\star)$, although
in usual slow-roll-inflation models, the spectral index fluctuation
$\delta n_s \propto \ln (k/k_\star ) dn_s/d\ln k (k_\star)$ is
typically restricted by $\vert 1-n_s\vert$. Table~\ref{tab:nrun}
summarizes the parameter values when running is allowed and
demonstrates that its inclusion has only a small effect on the other
parameters. For WMAP5 only, we find $dn_s/d\ln k
(k_\star)=-0.031^{+0.029}_{-0.028}$. The main tendency for the
negative value comes from the low-$\ell$ WMAP5 data. When the ACBAR
data is added, the median value is similar, $dn_s/d\ln k
(k_\star)=-0.037^{+0.023}_{-0.022}$, with a reduced error. The results
are nearly identical for CMBall, but more compatible with no-running
when LSS is added, $-0.016^{+0.019}_{-0.018}$.  The precise
measurement of the high-$\ell$ CMB power spectrum provided by ACBAR is
a potentially powerful constraint on running, but may also include
significant contributions from secondary anisotropies and foregrounds.
Therefore, in the last two columns we have included the effect of
marginalizing over the SZE or point source templates. In both cases,
the mean value of the running becomes more negative and more
significant at the ~2-sigma level.  A visual representation of the
impact of adding the ACBAR data is given in Fig.~\ref{fig:nrun} which
shows the correlation between $n_s$ and $dn_s/d\ln k$. The scalar
spectral index, $n_s(k_\star ) =0.918\pm 0.032$ for the WMAP5 + ACBAR
combination, depends on the choice of pivot point $k_\star$; a smaller
value would yield a higher result while a higher one would give an
even lower result.

\begin{deluxetable*}{c||ccccccc}

\tablecaption{Running Spectral Index Parameter Constraints }
\tabletypesize{\scriptsize}
\tablehead{ \colhead{}& \colhead{WMAP5} &\colhead{WMAP5+ACBAR} &\colhead{CMBall} &\colhead{CMBall+LSS} & \colhead{CMBall+$q_{\rm SZ}$}&\colhead{CMBall+SZ+$q_{\rm src}$}}

\startdata
$        \Omega_bh^2$  & $ 0.0219^{+0.0009}_{-0.0008}$ &$ 0.0219^{+0.0007}_{-0.0007}$ &$0.0220^{+0.0006}_{-0.0006}$ &$0.0225^{+0.0006}_{-0.0006} $ &$0.0217^{+0.0006}_{-0.0006}  $ &$0.0217^{+0.0007}_{-0.0006} $        \\
$        \Omega_ch^2$  & $ 0.117^{+0.009}_{-0.009}   $ &$ 0.119^{+0.008}_{-0.008}   $ &$0.120^{+0.008}_{-0.008}   $ &$0.109^{+0.005}_{-0.005}    $ &$0.122^{+0.008}_{-0.008}     $ &$0.124^{+0.008}_{-0.008}    $          \\
$          \theta$     & $ 1.040^{+0.003}_{-0.003}   $ &$ 1.041^{+0.003}_{-0.003}   $ &$1.042^{+0.002}_{-0.002}   $ &$1.042^{+0.002}_{-0.002}    $ &$1.041^{+0.003}_{-0.002}     $ &$1.041^{+0.002}_{-0.002}    $          \\
$            \tau$     & $ 0.090^{+0.008}_{-0.008}   $ &$ 0.091^{+0.009}_{-0.008}   $ &$0.092^{+0.009}_{-0.008}   $ &$0.091^{+0.009}_{-0.008}    $ &$0.092^{+0.009}_{-0.009}     $ &$0.094^{+0.009}_{-0.009}    $          \\
$             n_s$     & $ 0.923^{+0.043}_{-0.041}   $ &$ 0.918^{+0.033}_{-0.031}   $ &$0.914^{+0.030}_{-0.029}   $ &$0.948^{+0.026}_{-0.025}    $ &$0.894^{+0.031}_{-0.029}     $ &$0.896^{+0.030}_{-0.030}     $        \\
$    dn_s/d\ln(k)$     & $ -0.031^{+0.029}_{-0.028}  $ &$ -0.037^{+0.023}_{-0.022}  $ &$-0.039^{+0.021}_{-0.021}  $ &$-0.016^{+0.019}_{-0.018}   $ &$-0.052^{+0.022}_{-0.021}    $ &$-0.052^{+0.022}_{-0.022}    $       \\
$     {\cal A}_s$      & $ 3.09^{+0.05}_{-0.05}      $ &$ 3.10^{+0.05}_{-0.04}      $ &$3.10^{+0.05}_{-0.04}      $ &$3.07^{+0.04}_{-0.04}       $ &$3.10^{+0.05}_{-0.04}        $ &$3.11^{+0.05}_{-0.04}        $        \\
$  \Omega_\Lambda$     & $ 0.70^{+0.05}_{-0.06}      $ &$ 0.69^{+0.04}_{-0.05}      $ &$0.69^{+0.04}_{-0.05}      $ &$0.75^{+0.02}_{-0.03}       $ &$0.68^{+0.04}_{-0.05}        $ &$0.67^{+0.04}_{-0.05}        $        \\
$             Age$     & $ 13.8^{+ 0.2}_{- 0.2}      $ &$ 13.8^{+ 0.1}_{- 0.1}      $ &$13.8^{+ 0.1}_{- 0.1}      $ &$13.7^{+ 0.1}_{- 0.1}       $ &$13.8^{+ 0.1}_{- 0.1}        $ &$13.8^{+ 0.1}_{- 0.1}        $        \\
$        \Omega_m$     & $ 0.30^{+0.06}_{-0.05}      $ &$ 0.31^{+0.05}_{-0.04}      $ &$0.31^{+0.05}_{-0.04}      $ &$0.25^{+0.03}_{-0.02}       $ &$0.32^{+0.05}_{-0.04}        $ &$0.33^{+0.05}_{-0.04}        $        \\
$        \sigma_8$     & $ 0.81^{+0.04}_{-0.04}      $ &$ 0.83^{+0.03}_{-0.03}      $ &$0.83^{+0.03}_{-0.03}      $ &$0.79^{+0.03}_{-0.03}       $ &$0.83^{+0.03}_{-0.03}        $ &$0.84^{+0.03}_{-0.03}        $        \\
$          z_{re}$     & $ 11.7^{+ 1.7}_{- 1.7}      $ &$ 11.9^{+ 1.7}_{- 1.6}      $ &$12.0^{+ 1.7}_{- 1.6}      $ &$11.5^{+ 1.6}_{- 1.6}       $ &$12.2^{+ 1.8}_{- 1.7}        $ &$12.4^{+ 1.8}_{- 1.7}        $        \\
$             H_0$     & $ 68.4^{+ 4.2}_{- 3.8}      $ &$ 67.9^{+ 3.5}_{- 3.3}      $ &$67.7^{+ 3.3}_{- 3.2}      $ &$72.4^{+ 2.4}_{- 2.4}       $ &$66.7^{+ 3.3}_{- 3.1}        $ &$66.2^{+ 3.2}_{- 3.2}        $        \\
$     q_{\rm SZ}$ &\nodata&\nodata&\nodata&\nodata                                                                       &$0.77^{+0.12}_{-0.11}        $&\nodata       \\
$    q_{\rm src}$ &\nodata&\nodata&\nodata&\nodata&\nodata&                                                                                                                       $30^{+11}_{-29}  $       \\
$ \sigma_8^{q_{\rm SZ}}$ &\nodata&\nodata&\nodata&\nodata&                                                                $0.93^{+0.04}_{-0.05}       $&\nodata       \\
\enddata

\tablecomments{\small
Marginalized parameter constraints for models with a running of the spectral index. The tendency for the low-$\ell$ data to prefer negative values continues with the higher $\ell$ data, but at less than 2-sigma for CMB-only. The last two columns include marginalization over extra high-$\ell$ contributions from SZ and point sources. The basic parameters are stable with respect to this marginalization while the median value for $dn_s/d\ln k(k_\star)$ is increased slightly when the SZ or point source contribution is included. }

\label{tab:nrun}

\end{deluxetable*}


\subsection{Tensor modes}\label{sec:tensor}

We have run a limited number of cases including tensor modes,
characterizing their strength relative to scalar perturbations by $r$
and fixing the tensor tilt by $n_t \approx -r/8(1-r/16)$. The most
stringent upper limit for $r$ is given by the CMBall combination which
yields $r < 0.40$ (95\% confidence). This assumes a uniform prior
measure for $r$, as is conventional in parameter estimation, although
without much justification except that it is conservative. Adding LSS
tightens this limit. Our result can be compared with those obtained by
\citet{dunkley08}, $r<0.43$ (95\% confidence) for WMAP5 alone.
%

\section{Conclusions}\label{sec:conclusion}

We have used the complete ACBAR $150\,$GHz data set to measure the
CMB temperature anisotropy angular power spectrum.  Over
three seasons of observation, ACBAR dedicated 85K detector-hours to
CMB observations at 150 GHz and covered 1.7\% of the sky. The data are
calibrated by comparing CMB temperature maps for the largest ACBAR
fields with those produced by WMAP5.  This calibration is found to
be consistent with the previous planet-based and RCW38-based
calibrations, but with the temperature uncertainty reduced to
$2.05\%$. In the original preprint of this paper, the ACBAR band-powers 
were calibrated through comparison with WMAP3.
The new WMAP5 results became available during the review process and we 
have updated the ACBAR band-powers and cosmological parameters to take 
advantage of this new information. The ACBAR band-powers are otherwise
unchanged. The new WMAP5 parameters and calibration subtly changed the 
best fit WMAP5 and ACBAR model resulting in a decrease of the significance 
of the high-$\ell$ excess in the ACBAR data.

The ACBAR band-powers reported in Table~\ref{tab:bands} are the most
sensitive measurements to date of CMB temperature anisotropies for
multipoles between $\ell \sim 900$ and 3000. In this data, the fourth and fifth
acoustic peaks are significantly detected for the first time.
These precise measurements of the CMB temperature
anisotropies at high-$\ell$ are consistent with a spatially flat, dark
energy-dominated $\Lambda$CDM cosmology.  Including the
effects of CMB weak lensing in the computation of model power spectra improves 
the fits to the combined ACBAR+WMAP5 data.
The excellent fit of the $\Lambda$CDM cosmological model to
the combined ACBAR+WMAP5 data at $\ell \lesssim 2000$ is a strong
confirmation of the standard cosmological paradigm and gives us
confidence in the resulting parameter values.  The ACBAR data favor
higher median values of $\sigma_8$, $\Omega_m$, and $\theta$ than
those preferred by WMAP3; however, with the improved WMAP5 results for 
$\ell> 650$, this tension has been resolved. The
parameter values remain stable with the inclusion of additional CMB
and LSS data sets, with or without the marginalization over an SZE or point
source template for ACBAR. For example,  $\sigma_8\sim
0.80$ holds for all parameter variations in Table~\ref{tab:basic}, 
and even in Table~\ref{tab:nrun} with the inclusion of a running
spectral index.

We have performed strict jackknife tests with the data, and find that the
results are free of significant systematic errors. We have projected 
out templates derived from the FDS99 dust model and the PMN radio source
catalog from the ACBAR maps before estimating the band-powers and find
the residual contributions from these foregrounds to be negligible at
the current sensitivity.  The contribution of dusty
proto-galaxies is expected to be insignificant for all but the few
highest-$\ell$ band-powers, but remains poorly constrained due to our incomplete
knowledge of the spectral dependence of these sources.

Secondary anisotropies are expected to become important at small
angular scales. The ACBAR band-powers are slightly higher
(1.1$\sigma$) than expected for the primary CMB anisotropy at
multipoles above $\ell \sim 2000$.
We expect that some of this power results from contamination by dusty proto-galaxies. 
However, the combined signal is considerably smaller than the significant detection of 
excess power reported by the CBI experiment at $30\,$GHz.
A joint analysis of the CBI and ACBAR band-powers in the
multipole range of $2000\lesssim \ell \lesssim 3000$ 
argues strongly against the CBI excess having the spectrum of primary CMB 
anisotropy. 
These results are consistent with the excess power seen by CBI
being due to either the Sunyaev-Zel'dovich effect or radio source 
contamination.
Higher sensitivity observations over a broad range of frequencies are 
necessary in order to fully characterize CMB secondary anisotropies and eliminate 
potential foreground contamination.

\vspace{1cm}

The ACBAR program has been primarily supported by NSF office of polar
programs grants OPP-8920223 and OPP-0091840.  This research used
resources of the National Energy Research Scientific Computing Center,
which is supported by the Office of Science of the U.S. Department of
Energy under Contract No. DE-AC03-76SF00098. Some computations were
performed on the Canada Foundation for Innovation funded CITA
Sunnyvale cluster. Chao-Lin Kuo acknowledges support from a NASA
postdoctoral fellowship and Marcus Runyan acknowledges support from a
Fermi fellowship.  Christian Reichardt acknowledges support from a
National Science Foundation Graduate Research Fellowship. Some of the
results in this paper have been derived using the HEALPix
\citep{gorski05} package.  We thank members of the BOOMERANG team, in
particular Brendan Crill, Bill Jones, and Tom Montroy for providing
access to the B03 data, the pipeline used to generate simulation maps,
and assistance with its operation. We thank Antony Lewis for
discussions about ways to parameterize tests for weak lensing in the
data.

\appendix

\section{CALIBRATION }\label{app:calib}

The calibration used in K07 was linked to the Boomerang03 (B03)
calibration with observations of RCW38.  In this section, we describe
a new calibration using an $a_{lm}$-based comparison of CMB structure
observed by WMAP5 and ACBAR in 2005. This method is inspired by the
calibration scheme used to calibrate B03 to WMAP. The 2005 calibration
is carried to other years by an ACBAR-ACBAR power spectrum comparison
on fields observed in both years. The WMAP-ACBAR cross-calibration
method is described below, with a detailed accounting of uncertainty
in Table~\ref{tab:calerr}.

\paragraph{WMAP-ACBAR Calibration}

Calibrating with the CMB temperature anisotropies has two main
advantages. The first is that the calibration of the WMAP temperature
maps (at 0.2\% in temperature) is an order of magnitude more precise
than the flux calibration of the calibration sources ACBAR used in
previous releases. The second advantage is that the anisotropies have
the same spectrum as what is being calibrated, rendering the large
frequency gap between WMAP and ACBAR irrelevant.

\begin{figure*}
\epsscale{.9}
\plotone{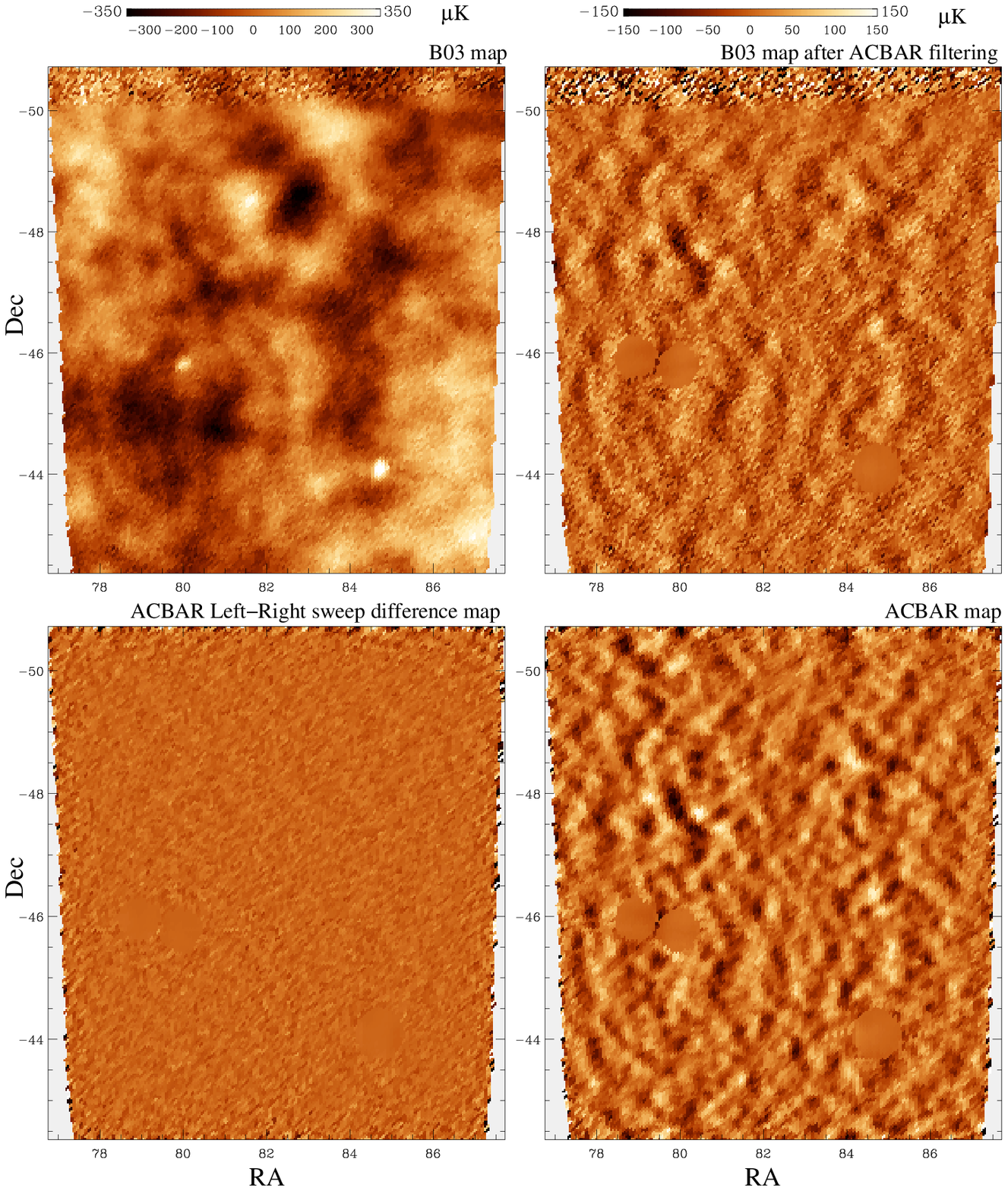}
\caption{ A comparison of observations of the CMB8 field made with B03
  and ACBAR. This field lies in the deep region of the B03 map. The
  top two maps are from B03.  The bottom two maps are from ACBAR. In
  the {\em top left} panel, the B03 map of the CMB8 field.  The
  dynamic range of this map is greater than that of the other three
  figures. The increased noise at one edge marks the edge of the B03
  deep coverage.  The ACBAR filtering is applied to the B03 map to
  create the map in the {\em top right} panel.  Directly below it in
  the {\em bottom right} panel is the ACBAR map of same region.  Note
  the clear correspondence between the CMB anisotropies observed by
  B03 and ACBAR. Three bright point sources have been masked. An ACBAR
  left-right sweep difference map is shown in the {\em bottom left}
  panel.  The power spectrum of this map (and the other 9 fields) is
  plotted in Fig. \ref{fig:sys}. }
\label{fig:cmb5}
\end{figure*}
 The two experiments have different scan patterns, noise, beam widths, and spatial 
filters that will effect the measured flux. In this analysis, we assume that the WMAP5 maps are effectively unfiltered except for the instrumental beam function. The two maps can then be represented as:
\[ S^{WMAP}_i = \int{T(x) B_{WMAP}(x_i - x)dx} +N^{WMAP}_i\]
\[ S^{ACBAR}_i = F_{ij} \int{T(x) B_{ACBAR}(x_j - x)dx} +N^{ACBAR}_i\]
where T is the underlying CMB signal, N is the instrumental noise, B is the beam function, and $F_{ij}$ is the ACBAR filter matrix as defined in Section \ref{sec:analysis}. We reduce the filtering differences by resampling the WMAP map using the ACBAR pointing information and applying the ACBAR spatial filtering to generate an `ACBAR-filtered' WMAP map. 
\[ S^{WMAP-equivalent}_i =  F_{ij} (\int{T(x) B_{WMAP}(x_j - x)dx} +N^{WMAP}_i)\]
The results of applying this algorithm to the B03 map is shown in Figure \ref{fig:cmb5}. We choose to do the absolute calibration via cross-power spectra rather than a direct pixel-to-pixel comparison of the maps.  Using cross-spectra significantly reduces the impact of the  noise model on the result. The significant beam differences between the experiments are more naturally dealt with in multipole space than in pixel space. We construct the ratio from the filtered maps:
\[ R = \Re\left( \left \langle \frac{a_{\ell m}^{WMAP-X *} * a_{\ell
        m}^{ACBAR-Z}}{a_{\ell m}^{ACBAR-Y *} * a_{\ell m}^{ACBAR-Z}
      (B_{\ell}^{WMAP-X}/B_{\ell}^{ACBAR})} \right \rangle \right) \]
where X can denote either the V- or W-band map for WMAP and Y/Z marks
either of two noise-independent ACBAR combinations.  There is a narrow
$\ell$-range from 256-512 useful for calibration. The range is limited
at high-$\ell$ by the rapidly falling WMAP beam function and at
low-$\ell$ by the ACBAR polynomial filtering which acts as a high-pass
filter. We choose to use the WMAP V \& W bands to take advantage of
their smaller beam size.

\begin{deluxetable}{lc}

\tablecaption{Error Budget for the $a_{\ell m}$-based ACBAR Calibration  }
\tabletypesize{\small}

\tablehead{ \colhead{Source} &  \colhead{Uncertainty (\%)}}

\startdata
Statistical Error in the Calibration ratio & 1.10 \\
\;\; $\ell$ dependence of the Calibration ratio & 1.1 \\
Statistical Error in the Transfer Function of the Calibration ratio & 0.35 \\
Uncertainty in the WMAP $B_{\ell}$ & 0.5 \\
Relative pointing uncertainty & 1.0 \\
Uncertainty in the Year-to-year ACBAR calibration & 0.3 \\
Uncertainty in the Transfer Function for the Power Spectrum & 0.5 \\
Contamination from foregrounds & 0.2 \\

\hline
WMAP5's Absolute Calibration  & 0.2 \\

Overall & 2.05\% \\ 
\enddata

\tablecomments{\small
The calibration of ACBAR using the WMAP5 temperature maps has multiple potential sources of error, tabulated here for reference.  
The dominant calibration uncertainties are due to noise in the WMAP maps at the angular scales used for calibration.  The uncertainty in the ACBAR beam function is comparable to the calibration uncertainty.}
\label{tab:calerr}
\end{deluxetable}


Monte Carlo simulations are used to determine the transfer function of
this estimator.  We generate CMB sky simulations convolved with the
respective instrumental beam functions using the
Healpix\footnote{http://healpix.jpl.nasa.gov} library. We resample
each realization and apply the ACBAR filtering matrix described above
to generate equivalent maps for each field. We expected and found a
small intrinsic bias as the beam convolution and filtering operations
do not commute: $B^{ACBAR} * F_{ij} B^{WMAP} \ne B^{WMAP} * F_{ij}
B^{ACBAR}$. We correct the real data by the $\ell$-dependent transfer
function measured in these simulations. The technique is easily
adapted to estimate the error caused by pointing uncertainties and to
confirm that the estimator is unbiased with the inclusion of
noise. The derived error in the transfer function is listed in
Table~\ref{tab:calerr}.

Foreground sources have the potential to systematically bias a
calibration bridging 60 to 150 GHz.  Radio sources, synchrotron
emission, dust, and free-free emission all have a distinctly different
spectral dependence than the CMB, which could lead to a calibration
bias. This risk is ameliorated by the positioning of the ACBAR fields
in regions of exceptionally low foregrounds.  Bright radio sources
detected in either experiment are masked and excluded from the
calibration. The calibration proved insensitive to the exact threshold
for source masking. We use the MEM foreground models in
\citet{hinshaw06} to estimate the RMS fluctuations of each foreground
relative to the CMB fluctuations and find that the free-free and
synchrotron fluctuations are less than 0.1\% of the CMB fluctuations
in all frequency bands while dust emission can reach 1.5\% of CMB
fluctuations in the 150 GHz maps. We test the effects of the most
significant foreground, dust, by adding the FDS99 dust model
\citep{finkbeiner99} to a set of CMB realizations. The resultant maps
are passed through a simulated pipeline as outlined in the previous
paragraph. We find that the addition of dust does not introduce a
detectable bias with an uncertainty of 0.2\%.

We perform a weighted average of the measured calibration ratio across all $\ell$-bin, field and band combinations after correcting for the estimated signal-only transfer 
functions. We estimate the calibration error to be 1.97\% for the 2005 data.
Table~\ref{tab:calerr} tabulates the contributing factors and error budget. 
We then propagate this $a_{\ell m}$-based calibration to the CMB observations 
done in 2001 and 2002.

\paragraph{ACBAR 2001-2002 and 2002-2005 Cross Calibrations}

We propagate the 2005 calibration into 2001 and 2002 by comparing the
2001 observations of the CMB2 field to the overlapping 2002 CMB4 field, and the 2002 observations of the CMB5 field to the 2005 observations of the CMB5 field. A power spectrum is 
calculated for each overlapping region and the ratio of the band-powers is used to derive a cross calibration. The procedures used are outlined in more detail in K07. We use the same relative calibration for 2001 as K07: $T_{2001}/T_{2002} = 1.238 \pm 0.067$.  We find cross-calibration factor for 2002 to be $T_{2005}/T_{2002} = 1.035 \pm 0.025$.  We apply these corrections to the data 
and determine the overall calibration uncertainty to be $2.05\%$ (in temperature units) based primarily on the uncertainties associated with WMAP/ACBAR-2005 cross calibration.

\bibliographystyle{apj}
\bibliography{merged}

\end{document}